%% file: ara.tex
\documentclass[10pt,journal,compsoc]{IEEEtran}

\usepackage{amsmath,amssymb,amsfonts}
\usepackage{graphicx}
\usepackage{xcolor}
\usepackage[hyphens]{url}
\usepackage[free-standing-units,per-mode=repeated-symbol,mode=text,detect-weight=true,detect-family=true]{siunitx}
\usepackage{pgfplots}
\usepackage{tikz}
\usepackage{booktabs}
\usepackage{glossaries}
\usepackage{cleveref}
\usepackage{pifont}
\usepackage[activate={true,nocompatibility},tracking=true,final]{microtype}
\usepackage{xspace}
\usepackage{ulem}
\usepackage{multirow}
\usepackage{threeparttable}
\usepackage[font=footnotesize]{caption}

\pgfplotsset{compat=1.18}

\newcommand\eew[2]{$\text{EW}_{#1}^{\text{#2}}$}
\newcommand\eg{e.g.,\xspace}
\newcommand\ie{i.e.,\xspace}
\newcommand\etal{et al.\xspace}

\newcommand{\ceil}[1]{\lceil #1 \rceil}
\newcommand{\cmark}{\ding{51}}
\newcommand{\xmark}{\ding{55}}

\DeclareSIUnit\bank{bank}
\DeclareSIUnit\cycle{cycle}
\DeclareSIQualifier\dp{DP}
\DeclareSIQualifier\sp{SP}
\DeclareSIQualifier\hp{HP}
\DeclareSIQualifier\bp{BP}
\DeclareSIQualifier\twop{2P}
\DeclareSIUnit\flop{FLOP}
\DeclareSIUnit\flops{FLOPS}
\DeclareSIUnit\gate{GE}
\DeclareSIUnit\lane{lane}
\DeclareSIUnit\op{OP}
\DeclareSIUnit\ops{OPS}
\DeclareSIUnit\request{request}
\DeclareSIUnit\core{core}
\DeclareSIUnit\macu{\gls{IPU}}
\DeclareSIUnit\pin{pin}
\DeclareSIUnit[quantity-product= ]\percent{\%}

\ifCLASSOPTIONcompsoc
  \usepackage[nocompress]{cite}
\else
  \usepackage{cite}
\fi

\ifCLASSINFOpdf
\else
\fi

\SetTracking{encoding={*}, shape=sc}{20}

\input{acronyms.tex}

\newif\ifshowrev
\newif\ifshowdel

\showrevfalse
\showdelfalse

\ifshowrev
  \ifshowdel
    \newcommand\revadd[1]{\textcolor{blue}{#1}\PackageWarning{}{#1!}}
    \newcommand\revdel[1]{\textcolor{red}{#1}\PackageWarning{}{#1!}}
    \newcommand\revmod[2]{\textcolor{red}{#1}\ \textcolor{blue}{#2}}
  \else
    \newcommand\revadd[1]{\textcolor{blue}{#1}\PackageWarning{}{#1!}}
    \newcommand\revdel[1]{}
    \newcommand\revmod[2]{\textcolor{blue}{#2}}
  \fi
\else
  \newcommand\revadd[1]{\textcolor{black}{#1}\PackageWarning{}{#1!}}
  \newcommand\revdel[1]{}
  \newcommand\revmod[2]{\textcolor{black}{#2}}
\fi

\begin{document}

\title{Ara2: Exploring Single- and Multi-Core Vector Processing with an Efficient \revmod{RVV1.0}{RVV 1.0} Compliant Open-Source Processor}

\author{Matteo~Perotti,~\IEEEmembership{Student Member,~IEEE,}
        Matheus~Cavalcante,~\IEEEmembership{Student Member,~IEEE,}
         Renzo~Andri,~\IEEEmembership{Member,~IEEE,}
          Lukas~Cavigelli,~\IEEEmembership{Member,~IEEE,}
        and~Luca~Benini,~\IEEEmembership{Fellow,~IEEE}%
  \IEEEcompsocitemizethanks{%
    \IEEEcompsocthanksitem{} Matteo~Perotti is with the
     Integrated Systems Laboratory (IIS), ETH Zurich, 8092 Zurich, Switzerland. E-mail:
     mperotti@iis.ee.ethz.ch%
    \IEEEcompsocthanksitem{} Matheus Cavalcante is with the
     Electrical Engineering Department, Stanford University, California 94305, USA. E-mail:
     mcv@stanford.edu%
    \IEEEcompsocthanksitem{} Renzo~Andri and Lukas Cavigelli are with the Huawei Zurich Research Center. E-mail: \{renzo.andri, lukas.cavigelli\}@huawei.com%
    \IEEEcompsocthanksitem{} Luca Benini is with the Integrated Systems
     Laboratory (IIS), ETH Zurich,
     Switzerland, and also with the Department of Electrical, Electronic and
     Information Engineering (DEI), University of Bologna, 40126 Bologna,
     Italy. E-mail: lbenini@iis.ee.ethz.ch.
  }%
  \thanks{\textcopyright\ 2024 IEEE.  Personal use of this material is permitted.  Permission from IEEE must be obtained for all other uses, in any current or future media, including reprinting/republishing this material for advertising or promotional purposes, creating new collective works, for resale or redistribution to servers or lists, or reuse of any copyrighted component of this work in other works.}
  \thanks{DOI: 10.1109/TC.2024.3388896}
}

\markboth{}%
{Shell \MakeLowercase{\textit{et al.}}: Bare Demo of IEEEtran.cls for Computer Society Journals}

\IEEEtitleabstractindextext{
\begin{abstract}
Vector processing is highly effective in boosting processor performance and efficiency for data-parallel workloads. In this paper, we present Ara2, the first fully open-source vector processor to support the RISC-V V 1.0 frozen ISA. We evaluate Ara2’s performance on a diverse set of data-parallel kernels for various problem sizes and vector-unit configurations, achieving an average functional-unit utilization of 95\% on the most computationally intensive kernels. We pinpoint performance boosters and bottlenecks, including the scalar core, memories, and vector architecture, providing insights into the main vector architecture’s performance drivers. Leveraging the openness of the design, we implement Ara2 in a 22nm technology, characterize its PPA metrics on various configurations (2-16 lanes), and analyze its microarchitecture and implementation bottlenecks. Ara2 achieves a state-of-the-art energy efficiency of 37.8 DP-GFLOPS/W (0.8V) and 1.35GHz of clock frequency (critical path: $\sim$40 FO4 gates). Finally, we explore the performance and energy-efficiency trade-offs of multi-core vector processors: we find that multiple vector cores help overcome the scalar core issue-rate bound that limits short-vector performance. For example, a cluster of eight 2-lane Ara2 (16 FPUs) achieves more than 3x better performance than a 16-lane single-core Ara2 (16 FPUs) when executing a 32x32x32 matrix multiplication, with 1.5x improved energy efficiency.
\end{abstract}

}

\bstctlcite{IEEE:BSTcontrol}

\maketitle

\IEEEdisplaynontitleabstractindextext

\IEEEpeerreviewmaketitle

\IEEEraisesectionheading{\section{Introduction}
\label{sec:intro}}

\IEEEPARstart{I}{n} 1976, the Cray-1 supercomputer~\cite{cray1978} achieved the world's top performance of \qty{250}{\mega\flops} 
by exploiting \gls{DLP} through a vector \gls{ISA}. 
Almost half a century later, in 2020, another vector architecture took the highest spot in the TOP500 list by scoring around \qty{537}{\peta\flops} of peak performance: the Fugaku supercomputer, based on Fujitsu's A64FX processor and Arm's \gls{SVE}.

Despite an increase in raw computer performance by nine orders of magnitude, today's computing systems struggle to meet the performance requirements of ubiquitous \gls{AI} and \gls{ML} workloads. 
The amount of data to process grows faster than the technology-driven processing-speed gain~\cite{fujitsu-comp-demand}, and the power required to sustain the ever-increasing performance needs is the main limiting factor from the cloud (operating cost, cooling) to the edge (battery lifetime, weight).
It comes as no surprise that the quest for energy efficiency is crucial for today's computing systems \cite{7079482} across all performance profiles, from supercomputers (Green500 ranking) to ultra-low power \gls{AI} processors (MLPerf Tiny).

While the Cray-1 vector processor was conceived for pure performance in the heydays of \gls{ECL} bipolar transistor technology, the recent feat of Fugaku proves that vector \glspl{ISA} are well-positioned to answer the conflicting needs of rapidly increasing computational demands under a tightly-constrained power budget. First, they effectively address the \gls{VNB} by reducing the number of instructions fetched from memory through the execution of the same instruction on multiple elements, in the so-called \gls{SIMD} execution model. 
Furthermore, vector processors leverage a \gls{VRF} to enhance data reuse in computationally intensive applications while remaining vector-length agnostic, as the vector length can be programmed at runtime. 

After the release of Arm's \gls{SVE} and SVE2 vector \glspl{ISA}, RISC-V also developed its \gls{RVV} vector extension, getting much interest due to its openness and extensibility. \gls{RVV} has reached frozen state with its 1.0 release in 2021 after six years of design and standardization efforts. 
Companies and academia developed numerous vector designs supporting various pre-release versions of \gls{RVV} and targeting different application domains such as \gls{HPC} \cite{Ara2020,Minervini2022,atrevido}, \gls{DNN} \cite{assir2021arrow}, \gls{AI}/\gls{AR} \cite{SiFiveX280}, real-time applications \cite{Platzer2021}, and \gls{IoT} endpoints \cite{Xuantie2020}.  
These designs are characterized by their heterogeneity in terms of deployed system architectures: \cite{Ara2020,Minervini2022,AndesNX27V,Platzer2021} are single-core systems, while \cite{Spatz2022,Xuantie2020,SiFiveP270,SiFiveX280} are multi-core or multi-core-ready. 
Among them, \cite{Ara2020,Platzer2021,Spatz2022} work with a relatively simple in-order core, while \cite{Minervini2022,AndesNX27V,Xuantie2020,SiFiveP270,SiFiveX280,SiFiveP870} deploy more aggressive superscalar or out-of-order scalar pipelines.

These designs show a wide spectrum of vector architectures, but there is a fundamental lack of detailed architectural performance studies on the effects that the design parameters of a vector processor (\gls{VRF} organization and size, number of \glspl{PE}, target application vector length) and system configurations (scalar core, scalar caches, single- or multi-core configuration) have on the system's performance and energy efficiency. As already noticed in \cite{Ara2020}, the scalar core can highly limit the system's performance when processing short vectors. 
Moreover, despite the efforts to propose standard \gls{RVV} benchmarks to ease architectural development and comparison \cite{ralc88}, the aforementioned publications provide performance data only for, in the best case, a small set of kernels.

In the constellation of RISC-V-based Vector processor architectures, Ara~\cite{Ara2020} was among the earliest ones. Its internal number of \glspl{FPU} could be scaled from 2 to 16, and the design, implemented in a 22nm technology, reached 1 GHz in typical conditions in all the configurations, with a leading \gls{SoA} of 41 DP-GFLOPS/W energy efficiency at its peak and 97\% of \glspl{FPU} peak utilization. The architecture, though, was compliant with only a limited set of \gls{RVV} 0.5, the very first and preliminary version of \gls{RVV}, missing key instructions like vector reductions, ubiquitous in the \gls{ML} domain.  
The \gls{ISA} architectural specifications and the software ecosystem deeply changed with the \gls{RVV} updates, and \revmod{RVV0.5}{RVV 0.5}-compliant designs are now obsolete. 

In this paper, we \revmod{present the \revmod{RVV-1.0}{RVV 1.0} Ara2 vector processor and leverage its efficient parametric architecture to explore a wide range of configurations to optimize computational efficiency}{extend our previous work \cite{Perotti2022} by presenting a thorough analysis of an RVV 1.0 vector processor's performance behavior over a wide range of kernels from different domains. We analyze its performance and scalability in a modern 22-nm technology, exploring the performance/power trade-off of multi-core vector architectures. We implemented support for all the RVV 1.0 instructions not supported in \cite{Perotti2022}, and we describe the micro-architectural optimizations that allow the architecture to scale up}.
To our knowledge, Ara2 is the first open-source vector processor to support the specification RISC-V V 1.0.
We analyze the frozen \gls{RVV} specification and its major effects on the microarchitecture of a vector processor, highlighting the novelties/changes and the issues that arise from them. 

Furthermore, Ara2 aims to close the gap between the various \gls{RVV} architectures proposed throughout the years and the lack of thorough performance/efficiency analyses for microarchitecture and implementation. 
The key contributions of this work are:

\begin{itemize}
    \item The Ara2 microarchitecture and design\footnote{https://github.com/pulp-platform/ara}, with support for the RISC-V ``V'' 1.0 \gls{ISA}. We analyze the changes from the previous preliminary \gls{RVV} specifications and discuss their impact on the system microarchitecture. To the best of our knowledge, Ara2 is the first open-source processor to support \gls{RVV} 1.0. 
    \item An in-depth analysis of performance depending on the application vector length on benchmark kernels from different application domains (Linear algebra, \gls{DSP}, \gls{ML}). Furthermore, we analyze the effect that the scalar processor and its memory system have on performance.
    \item An implementation of the Ara2 system in a 22nm \gls{FD-SOI} technology and its \gls{PPA} metric analysis from 2 to 16 lanes, with insights into the impact that microarchitectural optimizations of the most critical all-to-all unit (\gls{SLDU}) have on the scalability of the vector architecture.
    \item A performance/efficiency trade-off study on different hardware configurations, varying the number of vector cores and lanes per vector core on different problem sizes. We demonstrate how a multi-core vector architecture can mitigate the performance impact of the scalar core's issue-rate limitation and identify the most efficient architecture for specific application vector lengths.
\end{itemize}

The paper is structured as follows: Sections \ref{sec:intro} and \ref{sec:relworks} present an introduction and a background to contextualize the overall work. \revmod{\Cref{sec:riscvv} discusses the main changes between \gls{RVV} 0.5 and \gls{RVV} 1.0, and \Cref{sec:lanes} the implications on a lane-based vector processor, while, in}{In} \Cref{sec:arch}, we present the updated architecture\revmod{.
In}{, while in} \Cref{sec:experimentsetup}, we describe our experiment setup and\revdel{ describe} the benchmark kernels used for our evaluations. Then, we present performance analysis, physical implementation results, and a performance/efficiency trade-off study on single- and multi-core architectures in Sections \ref{sec:perf}, \ref{sec:phy}, and \ref{sec:multi}, respectively. Finally, we provide a comparison with the state-of-the-art in \Cref{sec:comp}.

\section{\revmod{Background and Related Work}{Related Work and Background}}
\label{sec:relworks}

\revdel{The vector processor architecture was first implemented by the Cray-1 supercomputer. }A vector processor is characterized by its internal \gls{VRF}, composed of a fixed number of registers whose size is upper-bounded by the \gls{ISA} (usually with a very loose upper bound). Arm \gls{SVE} and \gls{RVV} feature 32 vector registers, each of which with a size from 128 to 2048 (\gls{SVE}) or 64k (\gls{RVV}) bits. The \gls{VRF} can be conceptualized as L0 storage, meant to buffer data elements re-used multiple times close to the \glspl{PE} (with PE, we identify \revdel{every execution unit within a vector architecture, e.g.,} \gls{FPU}\revadd{s}, Multipliers, \glspl{ALU}, etc.)\revdel{ to avoid more costly memory accesses}. 
\revdel{Buffering more elements in the \gls{VRF} lowers the pressure on the data memory [Kung1986].
The \gls{VRF}, acting as an L0 buffer, also allows vector processors to be latency-tolerant with respect to memory stalls since every instruction occupies the \glspl{PE} for multiple cycles.}As noticed in \cite{vizcaino2023short}, \revmod{the tolerance against memory latency is generally higher}{vector processors are memory-latency tolerant, especially} when \revmod{the architecture works}{working} on longer vectors.
Moreover, the vector architecture amortizes the instructions' fetch, decode, and issue cost \revdel{of each vector instruction} by executing it on as many \revdel{data} elements as they fit in a vector register.\revdel{ Intuitively, the longer the vector, the less pressure on the instruction memory.}

\revdel{Two examples of successful commercial vector architectures are the NEC SX-Aurora TSUBASA and the Fujitsu A64FX.
The latest generation of SX-Aurora TSUBASA belongs to the NEC SX architecture family and features a 16-nm technology 1.6 GHz 16-core architecture equipped with a 128-KiB \gls{VRF} per core, composed of 64 vector registers connected to three 32-way \gls{FMA} pipelines, totaling a peak performance of 192 DP-FLOP/cycle per core and 4.91 TFLOPs for the whole system. 
Instead, the 7-nm FinFET-technology A64FX is an Arm-\gls{SVE}-based multi-core cluster-based architecture, where clusters of 13 cores (core-memory groups) are individually coupled with 8-GiB HBM2 modules and are connected together through a TofuD \gls{NoC} interconnect, with a peak theoretical performance of almost 3.38 TFLOPs.
All non-RISC-V vector architectures are commercial products and the microarchitectural details remain trade secrets. 
On the other hand, [] summarizes the \gls{SoA} of the RISC-V-based processors, which have been developed by companies as well as researchers. Ara2 is the first processor that supports \gls{RVV} 1.0 (Vicuna [] does not implement the floating-point and is only compliant with the Zve32x subset). }

\revdel{Among these vector architectures, [][][][][][], are multi-core or multi-core-ready, testifying the interest of industry towards coupling the two paradigms. 
[] provides a quantitative comparison of our design and the architectures in [].}

\begin{table}
\captionsetup{font=footnotesize}
\caption{Overview of RISC-V Vector Processors.}
\label{tab:relwork}
\scriptsize
\begin{threeparttable}
\begin{tabular}{@{}r@{\hspace{1mm}}l@{\hspace{1mm}}c@{\hspace{1mm}}c@{\hspace{1mm}}c@{\hspace{1mm}}c@{\hspace{1mm}}c@{\hspace{1mm}}c@{\hspace{1mm}}c@{\hspace{1mm}}c@{\hspace{1mm}}c@{}}
\toprule
 &\textbf{Core Name} & \textbf{RVV} & \textbf{Target} & \textbf{XLEN}  & \textbf{float} & \textbf{VLEN} & \textbf{Split VRF}  & \textbf{Open-}              \\
 & & \textbf{version} &  & \textbf{(bit)}    & \textbf{supp.} &  \textbf{(bit)} & \textbf{(lanes)}  & \textbf{Source}              \\ \midrule
 & This work & 1.0 & ASIC & 64 & \cmark & 1024\tnote{a} & \cmark & \cmark \\
\cite{SiFiveP870} & \textbf{SiFive P870} & 1.0 & ASIC & 64 & \cmark & 128 & \textbf{?} & \xmark \\ 
\cite{SiFiveX280} & \textbf{SiFive X280} & 1.0 & ASIC & 64 & \cmark & 512 & \textbf{?} & \xmark \\ 
\cite{SiFiveP270} & \textbf{SiFive P270} & 1.0rc & ASIC & 64 & \cmark & 256 & \textbf{?} & \xmark \\ 
\cite{AndesNX27V} & \textbf{Andes NX27V}  & 1.0  & ASIC & 64 & \cmark & 512 & \textbf{?} & \xmark \\ 
\cite{atrevido} & \textbf{S.D. VU} & 1.0 & ASIC & 64 & \cmark & 128-4096 & \cmark & \xmark \\ 
\cite{Spatz2022} & \textbf{Spatz} & 1.0\tnote{e} & ASIC & 32 & \xmark & 128-512 & \xmark & \xmark \\ 
\cite{Platzer2021} & \textbf{Vicuna}       & 0.10 & FPGA & 32 & \xmark & 128-2048 & \xmark & \cmark \\ 

\cite{assir2021arrow} & \textbf{Arrow}        & 0.9  & FPGA & 32 & \xmark  & \textbf{?} & \cmark \tnote{b} & \xmark \\ 

\cite{Johns2020} & \textbf{Johns \etal} & 0.8        & FPGA  & 32  & \xmark  & 32 & \xmark & \xmark \\ 

\cite{Minervini2022} & \textbf{Vitruvius+} & 0.7.1  & ASIC    & 64  & \cmark & 16384 &  \cmark & \xmark  \\
\cite{Xuantie2020} & \textbf{XuanTie 910} & 0.7.1  & ASIC    & 64  & \cmark & 128\tnote{a} &  \cmark & \xmark\tnote{c}  \\ 

\cite{risc-v-squared} & \textbf{RISC-}$\mathbf{V^2}$ & \textbf{?} & ASIC & \textbf{?} &  \xmark & 256 & \xmark & \cmark \tnote{d} \\ 

\cite{Ara2020} & \textbf{Ara} & 0.5       & ASIC    & 64  & \cmark & 4096 \tnote{a} &  \cmark & \cmark \\ 

\cite{hwachav5} & \textbf{Hwacha} & Non-Std. & ASIC       & 64  & \cmark & 512\tnote{a} &  \cmark & \cmark \\ 
\bottomrule
\end{tabular}
\begin{tablenotes}
\item[a] VLEN per lane. \item[b] VRF is split horizontally. \item[c] The vector unit is not open-source. \item[d] The scalar core is not open-source. \item[e] Limited subset of \gls{RVV}.
\end{tablenotes}
\end{threeparttable}
\end{table}

\revadd{\Cref{tab:relwork} summarizes recently published \gls{RVV} designs and compares them with Ara2. One key implementation challenge tied to the ratified \gls{RVV} 1.0 \gls{ISA} is matching a scalable lane-based architecture with the \gls{VRF} byte layout, mixed-width operations, and the \gls{VRF} writing policy. As detailed in \cite{Perotti2022}, Ara2 assigns consecutive elements to consecutive lanes to ease mixed-width operations. This byte layout exposes the following challenges:}

\textbf{Source Registers}
Every time a source register is read with an \revadd{Element Width} \eew{vs}{new} different than its previously encoded \eew{vs}{old}, \revmod{Ara2 should}{the hardware has to} reshuffle its bytes \revadd{across the lanes} to reinterpret its content according to \revmod{the new element width}{\eew{vs}{new}}. \revdel{Generally, if the operation is executed outside of the lanes, the units can automatically reshuffle the data elements on the fly.}
\revmod{In the case of in-lane operations, the \gls{SLDU} should reshuffle the source register content since, with a different \gls{EW}, bytes belonging to the same element can be in different lanes. We do this by injecting a reshuffle operation before the “offending” instruction.}{To do so, the hardware injects a reshuffle micro-operation before executing the instruction with \eew{vs}{new}. In Ara2, the reshuffle is performed by the Slide Unit (SLDU)}. 

\textbf{Destination Registers}
\gls{RVV} also mandates to support a tail-undisturbed policy. When in place, the destination elements past the last active one should not be modified.
When an \eew{vd}{new}-instruction writes a vector register \texttt{vd} previously encoded with \eew{vd}{old}$\ne$ \eew{vd}{new} and the old content of \texttt{vd} is not fully overwritten, the unmodified bytes \revadd{would} get corrupted\revdel{ since the byte mapping of \texttt{vd} is no longer unique}.

To avoid corrupting the tail elements, \revdel{when this situation occurs, }Ara2 \revmod{should}{has to} \revadd{match the element width encoding with \eew{vd}{new}. To do so, it} deshuffle\revadd{s} the destination register using \eew{vd}{old} and reshuffle\revadd{s} it back using \eew{vd}{new} by injecting a reshuffle operation (i.e., a vector slide with null stride) into the \gls{SLDU} before the offending instruction. 
Since it is not possible to know how many bytes need to be reshuffled unless both the vector length and the element widths are dynamically tracked for each vector register, the reshuffle acts on the whole register. No reshuffle is injected when the offending instruction writes a whole vector register\revmod{,}{.} 
\revmod{as}{It is important to note that} reshuffling hurts \revmod{the \gls{IPC}}{performance} if its latency cannot be hidden and if \revmod{this operation causes structural hazards on \revadd{the} following slide or reduction instructions.}{it precedes slide or reduction instructions, which would suffer from a structural hazard.}
\revmod{Reshuffling is costly, as the offending instruction has a \gls{RAW} dependency on the reshuffle.
The compiler can alleviate the problem by assigning \gls{EW}-encoded vector registers to instructions with the same \gls{EW}.}{The performance-aware programmer should try to avoid reshuffling by always writing the vector registers with the same \gls{EW}, \eg by using different vector registers to keep vectors with different \glspl{EW} when executing multiple mixed-width operations.}

\begin{figure}
    \centering
    \includegraphics[scale=0.3]{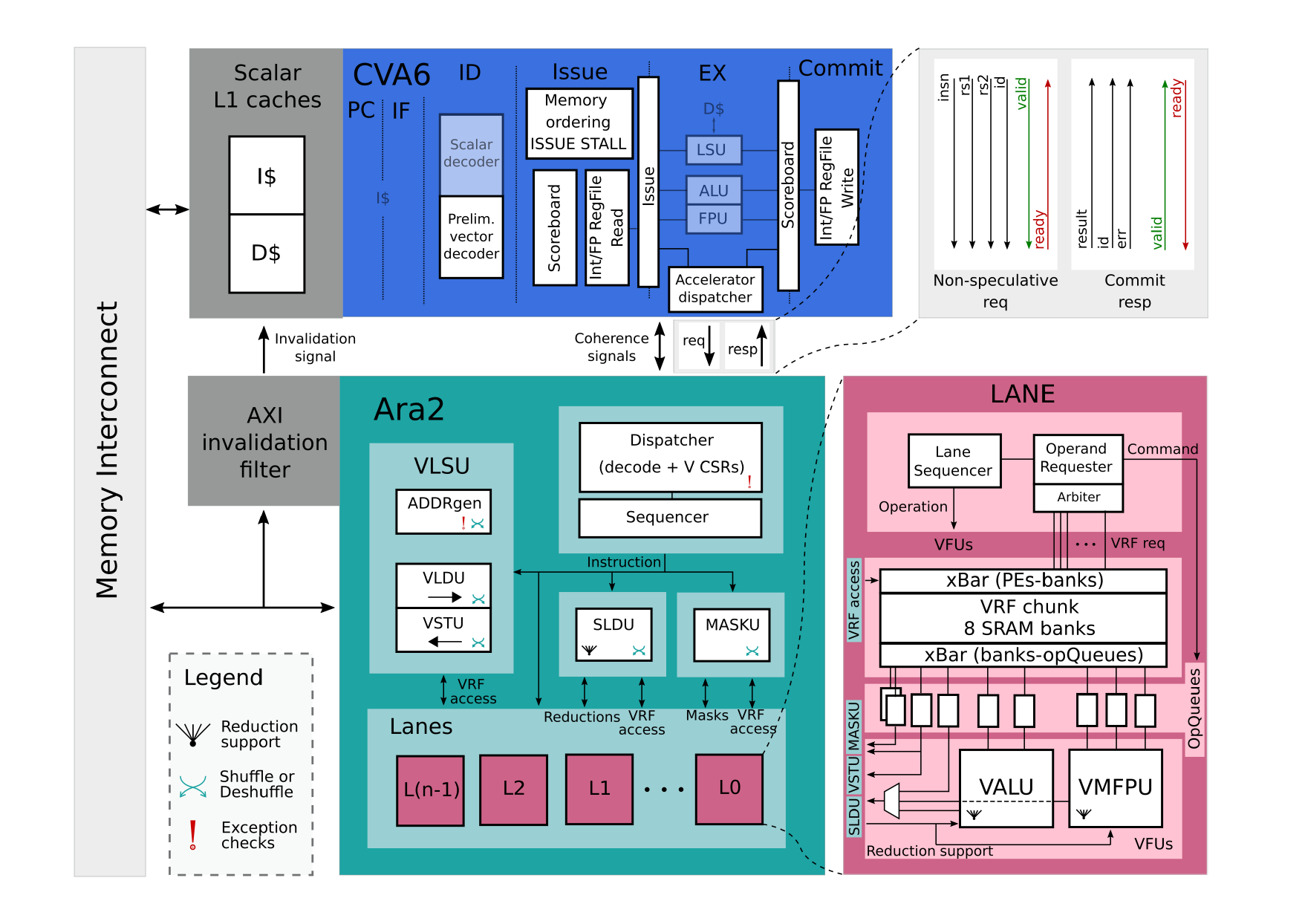}
    \caption{Top-Level block diagram of the Ara2 system with the vector co-processor (green), a more detailed diagram of the lane (magenta), and the host scalar core CVA6 (blue).}
    \label{fig:main-dg}
\end{figure}

\section{Architecture}
\label{sec:arch}

\revadd{To achieve compliance with \gls{RVV} 1.0, we added to \cite{Perotti2022} the support for fixed-point arithmetic, floating-point reductions, scalar element moves, missing mask instructions, indexed permutations, and segmented memory operations}.
In \Cref{fig:main-dg}, we show the block diagram of the scalar core CVA6 and the vector unit Ara2.
The architecture is based on Ara \cite{Ara2020}, with the following major enhancements:

\textit{Decoding: }
With the new RISC-V V specifications, the encoding of the vector instructions now fully specifies the data type of the vector elements on which the instruction operates. This allows moving most of the decoding logic and vector-specific \glspl{CSR} from CVA6 to the vector unit\revdel{, making CVA6 more agnostic on the V extension}.
In the updated architecture, CVA6 keeps only the pre-decoding logic strictly needed to know 1) if the instruction is a vector instruction (to dispatch it to the vector unit when it reaches the head of the scoreboard), 2) if the vector instruction is a memory operation (to enforce cache coherence), and 3) if the instruction needs a scalar value from the integer or floating-point scalar register files.

\revadd{The full decoding of the vector instructions is done in Ara2's dispatcher, which also contains the vector \glspl{CSR} and the byte layout encoding of each vector register. The dispatcher checks if the decoded instruction is legal and, if needed, injects reshuffle instructions. This is the only place in which the LMUL parameter plays an important role, as each lane's operand requester sees the \gls{VRF} as a contiguous memory region from which fetching bytes. Higher LMUL values will allow for higher vector lengths and, when reading/writing the \gls{VRF}, access more contiguous bytes.
At decoding time, entire sets of instructions are aliased with the same decoded operation. The dispatcher automatically aliases the whole register moves with regular vector moves, modifying the vector length to always move a whole register unconditionally. The architecture always uses the undisturbed tail policy.}

\textit{CVA6-Vector Unit Interface: }
While decoding, CVA6 identifies vector instructions, pushes them to a dispatcher queue, and dispatches them to the accelerator once they are no longer speculative. \revadd{This helps the vector architecture avoid expensive roll-back logic in case of mis-speculation. As shown in \Cref{sec:perf}, when dealing with medium-long vectors, waiting until the instruction is known to be non-speculative has a negligible impact on performance, as each instruction keeps Ara2 busy for several cycles.}

The interface between the host processor CVA6 and the vector unit is generalized: the unit is implemented as a modular accelerator with its own CSR file. 
\revadd{The interface can be divided into an instruction and a memory interface. The former is a valid-ready request-response interface. CVA6 forwards the vector instruction and a maximum of two 64-bit operands fetched from the integer or floating-point scalar register file, together with an identifier. Ara2 answers with a 64-bit bus for the result and information about the occurred exception when needed.}
\revadd{The memory interface is composed of control signals that help enforce memory ordering.}

\textit{Memory \revadd{Ordering and} Coherence: }
CVA6 and the vector unit have separate memory ports, and CVA6 has a private L1 data cache.
At the same time, the \mbox{RISC-V} ISA mandates a strictly coherent memory view between the scalar and vector processors.
Ara~\cite{Ara2020} violates this requirement and needs explicit memory fences that write back and invalidate the CVA6 data cache between accesses on shared memory regions, adding a significant performance overhead and reducing code portability. In Ara2, we develop a hardware mechanism to ensure coherence. We adapt the CVA6 L1 data cache to a write-through policy so that the main memory, which is accessed by the vector unit as well, is always up-to-date. When the vector unit performs a vector store, it invalidates the corresponding cache lines in the CVA6 data cache. Moreover, we issue 1) scalar loads only if no vector stores are in-flight, 2) scalar stores only if no vector loads or stores are in-flight, and 3) vector loads or stores only if no scalar stores are pending.
\revadd{CVA6 keeps track of the issued loads and stores through dedicated counters. For example, the vector store counter is increased when a vector store is forwarded to the vector unit and decreased by the vector unit every time a vector store is over. The memory interface comprises signals to inform the scalar core that one vector load or store is over and if there are pending stores. Thanks to this mechanism, no memory disambiguation logic is needed between vector memory operations and scalar stores (and vice versa) since scalar/vector memory stores cannot collide with other vector/scalar memory operations.}

\revadd{\textit{Segmented Memory Operations: } Segmented memory operations can require non-negligible control logic and connections when optimized for all the possible data widths and segment values in a parametric lane-based architecture. Each Ara2 lane can send one 64-bit element per lane every cycle. During a segmented memory operation, the memory bus often contains more elements that map to the same lane. Thus, we decided not to optimize these operations and load/store only one element per cycle to simplify control and datapath.}

\textit{Mask Unit: }
After the update to RVV 1.0, the mask bits used by one lane can be stored in a different one. Thus, we designed a Mask Unit, which can access all the lanes at once to fetch, deshuffle at bit granularity, and dispatch the correct mask bits to each lane.
Moreover, the \gls{MASKU} combines the results of many vector mask instructions, such as \texttt{vcpop} and \texttt{vfirst}.
The introduction of an additional unit that accesses all the lanes leads to a greater routing complexity, especially when scaling up the number of lanes, as already noticed in~\cite{Ara2020}.

\begin{figure}
    \centering
    \includegraphics[width=\linewidth]{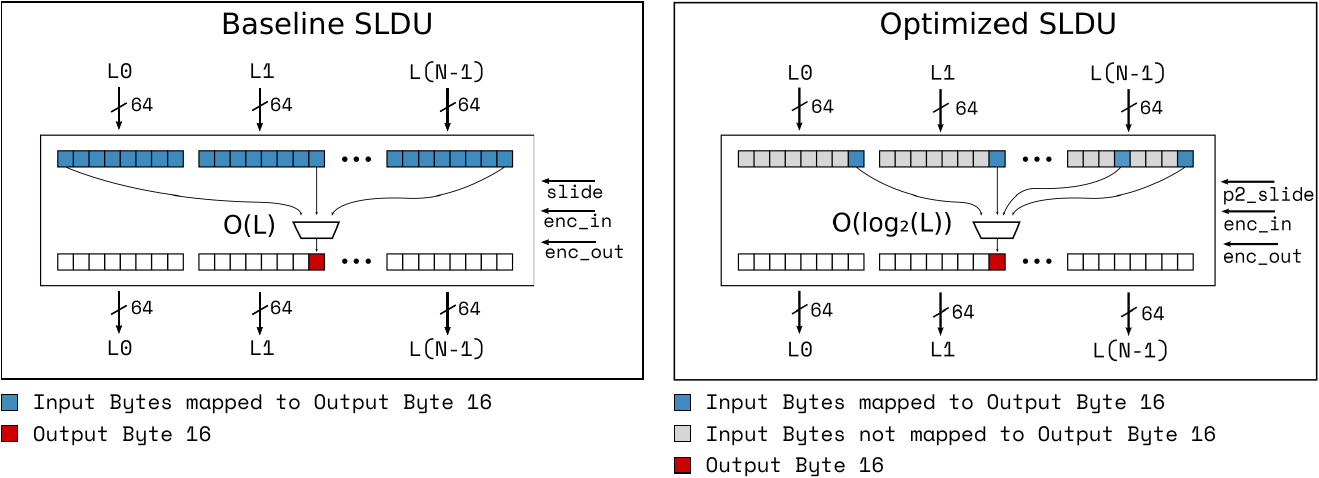}
    \caption{Comparison between the baseline and the optimized slide units, with a focus on how an arbitrary byte (15th) is mapped to the input bytes. The baseline slide unit supports arbitrary slide amounts and can slide and re-encode a vector in the same cycle. Each output byte is mapped to every input byte, so that the total number of connections is O($L^2$). In the optimized design, we support only power-of-two slides. Moreover, slides and re-encoding cannot happen at the same time. The total number of connections grows following an O($L \times log_2(L)$) behavior.}
    \label{fig:sldu}
\end{figure}

\begin{figure}
    \centering
    \includegraphics[width=\linewidth]{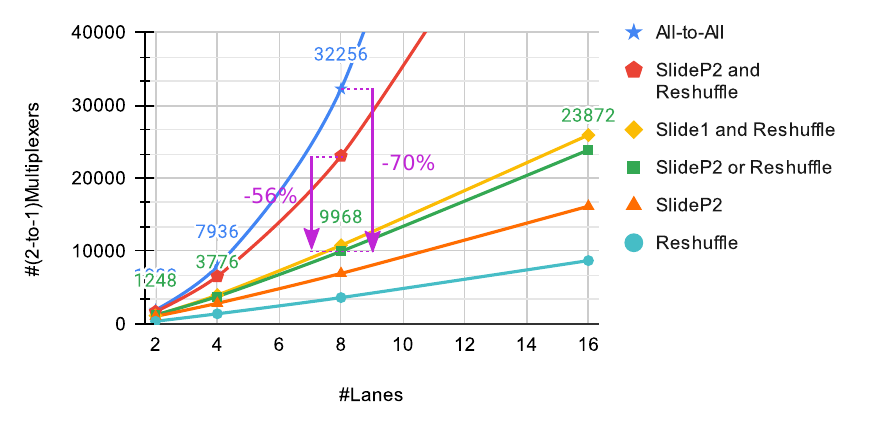}
    \vspace{-5mm}\caption{Number of 2-to-1 multiplexers needed to implement a slide unit as a function of the number of lanes with different configurations. An all-to-all slide unit connects every input byte to every output byte and supports every slide and reshuffle operation in one cycle. Other configurations support only slides by powers of two (SlideP2), slides by one (Slide1), and can slide and reshuffle or either slide or reshuffle.}
    \label{fig:sldu_conn}
\end{figure}

\textit{Optimized Slide Unit: }
The mask unit makes the physical implementation effort harder due to its all-to-all connections to/from the lanes. Reducing the complexity of each all-to-all connected unit is paramount to ease the physical implementation (placement and routing) of the system.
A preliminary study of how the area of the system scales with the number of lanes shows that the most problematic block is the slide unit, being the largest all-to-all connected unit and scaling quadratically with the number of lanes. This is expected, as the block supports slides by arbitrary amounts for element widths down to 8 bits by mapping the $8 \times L$ input bytes to the $8 \times L$ output bytes. 
This interconnect can be simplified by limiting the type of operations supported by the unit. For example, we can constrain the datapath to support only slide-by-1 or slide-by-power-of-two slide instructions, with the other strides handled with multiple micro-operations. 

In \Cref{fig:sldu_conn}, we attach a plot showing the number of 2-to-1 multiplexers needed to implement the interconnect with different slide unit configurations. This number is also a lower bound to the number of wires that connect the input bits before the slide to the output slided bits and is a valid metric to estimate the area of the interconnect and the wiring complexity during the routing stage. 
Since non-power-of-two slides and reshuffling are uncommon, we limit the strides fully supported in hardware to powers of two and handle the remaining strides with micro-operations\revadd{\ (\ie by decomposing the non-power-of-two-stride slide instruction in power-of-two-stride slide operations)}, saving up to ~70\% of the estimated area and wires, as shown in \Cref{fig:sldu}. 

\textit{Reductions}
Since our design has lanes, we implement reductions using a 3-steps approach: intra-lane, inter-lane, and SIMD reduction steps. The intra-lane reduction fully exploits the data locality within each lane, maximizing the ALU (or \gls{FPU}) utilization and efficiency, reducing all the elements already present in the lane. The inter-lane reduction moves and reduces data among the lanes in $\log_2(\ell)+1$ steps, $\ell$ being the number of lanes, using the slide unit; since there is a dependency feedback between the slide and ALU (or \gls{FPU}) units, the latency overhead of the communication is paid at every step. Finally, the \gls{SIMD} reduction reduces the \gls{SIMD} word, if needed; therefore, its latency logarithmically depends on the element width.

Floating-point reductions need special care since the \gls{FPU} is pipelined. To keep the throughput at its maximum, the internal pipeline registers of the \gls{FPU} are used as accumulators. 
During the intra-lane phase and after adding the first two elements together, each FPU is constantly fed with one operand per cycle. The other FPU input is either a neutral value (zero for the addition) or one of the partial sums coming from the FPU output, when available. 
In the best case, after $N/L$ cycles, each FPU pipeline register contains a partial sum, and the reduction will be completed in $R \times (1 + \log_2\ceil{R}) - (\ceil{R} - R) - 1$ cycles. Here, $N$ is the number of 64-bit packets of operands in the vector to reduce, $L$ is the number of lanes, and $R$ is the number of FPU pipeline stages. When $R$ is a power of two, the formula simplifies to $R \times (1 + \log_2(R)) - 1$.
Then, during both the inter-lane and SIMD reductions, the FPU latency directly affects each micro-operation. Since the number of FPU pipeline registers increases with the \gls{EW}, reductions on lower bit widths can be faster than the ones on higher bit widths.

\section{Experiment Setup}
\label{sec:experimentsetup}
In the following section, we describe our experiments in detail, including the system configuration, tools, and metrics used for performance characterization, physical implementation, and single-/multi-core trade-off evaluations. Then, we outline the details of our benchmark pools and the rationale behind selecting our kernels. In the following, we will use the term \textit{system} to refer to the system-under-test composed of CVA6, its caches\,\footnote{I\$: 4 KiB, 4 sets, 128-bit line. D\$: 8 KiB, 4 sets, 256-bit line}, and Ara2. Ara2 and the L1 caches connect to \revmod{a 1-cycle-latency}{an} SRAM memory\,\footnote{Memory: 2M words, $4 \times L$ Byte/word, $L$ being the number of lanes} via the AXI memory interconnect shown in \Cref{fig:main-dg}. \revadd{The latency from a memory request to a response is seven clock cycles for Ara2 and five for CVA6.}
While analyzing a complex memory hierarchy is out of the scope of this work, this simple configuration includes L1 caches and a standard system bus initiator with realistic latency. \revadd{This configuration enables us to account for the latency of the next level of the memory hierarchy while focusing on the characterization of the vector processor.}

\textbf{Performance analysis}: 
We characterize the system's performance in terms of \textit{raw throughput ideality}, which identifies the ratio between the measured raw throughput and the ideal maximum raw throughput achievable by the vector processor for a particular kernel\revadd{, only limited by its maximum performance or its memory bandwidth}. With \textit{raw throughput}, we identify the ratio between the number of “useful” operations of an algorithm and the number of clock cycles spent to execute it, counted from the first vector instruction dispatched from CVA6 until the last vector instruction fully executed by Ara2 (\textit{raw} since there is no frequency information). 
This metric helps study the ideality of the design on different vector lengths (different problem sizes). \revadd{Our experiment extends the performance analysis in \cite{Perotti2022}, which only focuses on two benchmarks run on a 4-lane Ara architecture}.
For our measurements, we run all benchmarks on our vector system using a cycle-accurate simulation, with all the benchmark instructions and data preloaded in the SRAM main memory.
\revadd{We compile our code with \texttt{-O3} optimizations and use assembly vector instructions or intrinsics within the vector kernels to decouple our experiment from the current RISC-V vector compiler technology and focus on the quality of the hardware}.

We measure the performance of each benchmark for different vector lengths (from 32 to 1024 Byte) and hardware configurations (Ara2 with 2, 4, 8, and 16 lanes) to constitute a baseline for further performance analyses.

\textbf{What-if analyses:}
We repeat the performance measurements by replacing \texttt{CVA6} and its L1-caches with a FIFO that contains the whole dynamic vector instruction trace of the program in analysis and the correct scalar register file entries to be forwarded to Ara2 if needed. 
This allows us to determine the impact of the scalar part of the system on the overall computation and set an upper bound on performance gains achievable by streamlining the scalar sub-system. In these conditions, performance is only limited by the vector co-processor.
Additionally, we track the number of L1-cache misses to study the effect of scalar memory system non-idealities on kernel performance. 
Finally, we further streamline the vector processor with modifications that would impact the vector processor hardware (e.g., by aggressively upsizing the internal queues and scoreboard) and study their impact on the system's performance.
\revmod{The overall experiment highlights}{If in \cite{Perotti2022} we only showed that the scalar memory system and its configuration have an influence on matrix multiplication's performance, the new set of experiments quantitatively highlights} the bottlenecks that limit performance\revadd{, their cause (scalar core, caches, or vector architecture),} and \revmod{the cost to overcome them}{ the performance effect of the Barber's Pole \gls{VRF} layout.}

\textbf{Physical implementation analysis}: 
After synthesizing, placing, and routing the system in 22nm FD-SOI technology using industrial-grade tools, we evaluate its power, performance, and area (PPA) metrics. The system includes CVA6, its L1 caches, the invalidation filter, Ara2, and the interconnects up to the system crossbar. To analyze the switching activity of the netlist internal signals, we use a cycle-accurate simulator to generate a VCD file with delay back-annotations from the physical place-and-routed design. \revmod{We repeat the analyses for systems with 2, 4, 8, and 16 lanes}{We implement the full Ara2 system with 2, 4, 8, and 16 lanes to study its scaling behavior and intrinsic limitations}. The input matrices \revadd{for power simulations} \revmod{are}{contain} samples from a uniform distribution over [0,1).

\textbf{Multi-Core analysis}: 
\revadd{We then extend our analysis to study how the vector core architecture behaves in a multi-core environment and explore its performance/efficiency trade-off.} The system is modified to accommodate multiple clusters of the previously implemented system, and the SRAM main memory is multi-banked to support multiple Ara2 instances, with one bank per Ara2 and every bank with a parallelism of $4 \times L$ Byte each, with $L$ being the number of lanes of each Ara2. Moreover,
we implement a lightweight synchronization engine via
system-level CSRs to let the clusters synchronize at the end of the execution. The main objective is to evaluate the impact of the multi-core vector architecture on performance and energy efficiency.

To do this, we vary the number of cores per cluster, the number of lanes of each vector core, and the application vector length of the floating-point matrix multiplication algorithm. The modified kernel used in this phase is slightly different from the one used in \Cref{sec:perf}, and scalar caches are not warmed up to ease the automatic VCD dumping script. However, the comparison is still fair as all the configurations run the same program.

We insert a synchronization point before and after the kernel execution, measure performance and energy efficiency within this time span, and simulate \texttt{fmatmul} between square matrices with different sizes ranging from 4x4 to 256x256 elements on systems with \{1, 2, 4, 8\} cores and \{2, 4, 8, 16\} lanes, with no more than 16 \glspl{FPU} in total.

\textbf{Benchmark selection:}\label{ssec:bmarks} \Cref{tab:bmarks} summarizes the benchmarks used to evaluate the Ara2 System's performance. The benchmarks include \revdel{a mix of }kernels from various domains, including signal processing, linear algebra\revdel{ (BLAS)}, and machine learning (ML). The kernels were chosen to stimulate all the main units of Ara2 and include both compute-bound and memory-bound kernels. Here is a brief overview of each benchmark:

\texttt{matmul} and \texttt{conv2d} are two compute-bound kernels \revdel{that are }used in multiple domains (Signal processing, linear algebra\revdel{(BLAS)}, \gls{ML}). Their performance behavior is crucial due to the kernels' ubiquity. We re-used \texttt{matmul} from \cite{Ara2020} and optimized the 3x7x7-kernel \texttt{conv2d} by keeping, after the initial setup steps, seven output vectors in the \gls{VRF} for every loaded input vector, to maximize data reuse. 
\texttt{fft} and \texttt{dwt} are well-known algorithms used in the signal processing domain. We implement a vectorized version of \texttt{fft} by exploiting the large Ara2 \gls{VRF} to buffer all the input samples. This avoids costly transfers to and from memory, in a similar way as shown in Bertaccini et al.\cite{9516654}, and sets a limit to $128 \times \mathrm{Lanes}$ input samples (e.g., 2048 input samples with 16 Lanes). We take \texttt{pathfinder}, \texttt{jacobi2d}, and \texttt{exp} from the RiVec benchmark suite repository \cite{ralc88github,ralc88}\revdel{, and optimize them with assembly instructions.} \revdel{We chose these three kernels from RiVec} since they were the easiest ones to adapt to bare-metal C code, as the suite was primarily developed to run on gem5 with OS support. \revadd{We tuned these three kernels to Ara2 by preloading scalar coefficients in advance and exploiting Ara2's large \gls{VRF} to buffer the vectors longer with the aid of assembly instructions}.
\revmod{We}{Finally, we} vectorized \texttt{dropout}, \texttt{roi-align}, and \texttt{softmax}, three kernels used in the \gls{ML} domain \revmod{, and}{to reduce neural network overfitting, extract a compact feature map from individual regions of interest in detection tasks, and calculate the final attention score at the end of a neural network or transformer model, respectively. Also,} we included \texttt{dotproduct} since it does not expose multiple dimensions for parallelization and forces \revmod{the implementation to use}{using} a reduction.

\begin{table}[t]
\captionsetup{font=footnotesize}
  \caption{Benchmarks used in the performance analysis.}
  \label{tab:bmarks}
  \centering
  \resizebox{\columnwidth}{!}{
  \setlength{\tabcolsep}{2pt}
  \begin{tabular}{lrrrrrrrrr}
    \toprule
    Program & Domain & DType & CB & M & S & SMO & IMO & R & Max Perf. [\revdel{FL}OP/cycle] \\\midrule
    \textbf{matmul} & Linear Algebra, ML & FP64 & Y & N & N & N & N & N & $\mathrm{1 \times 2.0 \times L}$ \\
    \textbf{conv2d} & Signal Processing, ML & FP64 & Y & N & Y & N & N & N & $\mathrm{1 \times 2.0 \times L}$ \\
    \textbf{dotproduct} & Linear Algebra & FP64, int64 & N & N & N & N & N & Y & $\mathrm{1 \times 0.5 \times L}$ \\
    \textbf{jacobi2d} & Stencil & FP64 & Y & N & Y & N & N & N & $\mathrm{1 \times 1.0 \times L}$ \\
    \textbf{dropout} & ML & FP32 & N & Y & N & N & N & N & $\mathrm{2 \times 0.25 \times L}$  \\
    \textbf{fft} & Signal Processing & FP32 & Y & Y & Y & N & Y & N & $\mathrm{2 \times 5/4 \times L}$  \\
    \textbf{dwt} & Signal Processing & FP32 & N & N & N & Y & N & N & $\mathrm{2 \times 0.5 \times L}$  \\
    \textbf{pathfinder} & Routing Algorithm & int32 & \revmod{N}{Y} & Y & N & N & N & N & $\mathrm{2 \times 1.0 \times L}$ \\
    \textbf{exp} & Scientific, ML & FP64 & Y & Y & N & N & N & N & \revdel{$\mathrm{1 \times 28/21 \times L}$}\revadd{$\mathrm{1 \times 30/23 \times L}$} \\
    \textbf{softmax} & ML & FP32 & Y & N & N & N & N & Y & \revdel{$\mathrm{2 \times 32/25 \times L}$}\revadd{$\mathrm{2 \times 34/27 \times L}$}  \\
    \textbf{roi-align} & ML & FP32 & N & N & N & N & N & N & \revdel{$\mathrm{2 \times 3/5 \times L}$}\revadd{$1 \times \mathrm{9/5 \times L}$} \\
    \bottomrule
    \multicolumn{10}{l} {\parbox[t]{1.35\columnwidth}{\textbf{CB}: Compute Bound, \textbf{M}: Masks, \textbf{S}: Slides, \textbf{SMO}: Strided Mem. Ops., \textbf{IMO}: Indexed Mem. Ops., \textbf{R}: Reductions 
    }}  \\
  \end{tabular}
  }
\end{table}

\section{Performance characterization}
\label{sec:perf}
In this section, we analyze the system's performance and how it changes with the system configuration, the scalar core, the configuration of the scalar memory hierarchy, the number of lanes, and the application vector length.

\subsection{Lanes and Vector Length}
\label{ssec:lanes_vec_len}
\begin{figure}
    \centering
    \includegraphics[width=\linewidth]{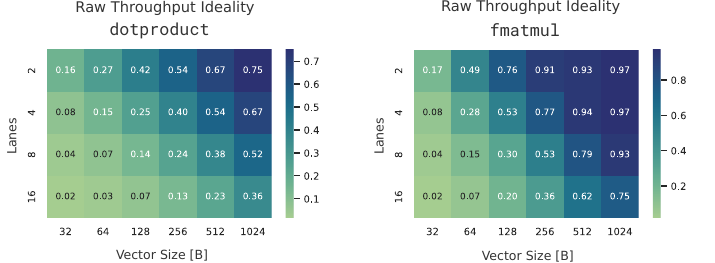}
    \caption{Correlation between Raw throughput Ideality and Byte/lane ratio for dotproduct (\revmod{top}{left}) and fmatmul (\revmod{bottom}{right}). The raw throughput ideality tends to be similar when the number of elements per lane is the same (on the diagonals).
    }
    \label{fig:elm_per_lane}
\end{figure}

When we scale up the Ara2 co-processor by doubling its number of lanes, we also double 1) its AXI bus data width and 2) the parallelism of the all-to-all internal units (\gls{MASKU}, \gls{SLDU}, \gls{VLSU}). For a deeper understanding of the performance scalability, \Cref{fig:elm_per_lane} shows the correlation between the performance ideality of the Ara2 co-processor, its number of lanes, and the application vector size, for two different kernels: \texttt{dotproduct} and \texttt{fmatmul}. Specifically, on the $x$-axis of the plots, we list different vector sizes (in Byte), \ie the number of vector elements times the element size. For example, the entry (\texttt{8L}, \texttt{512 Byte}) corresponds to the performance ideality achieved by an 8-lane system when dealing with vectors composed of 64 64-bit elements each. When the ratio \#Byte/\#Lane is constant, the achieved performance ideality is similar for systems with different configurations. This means that the raw throughput almost doubles when the number of lanes doubles if the number of elements per lane is constant. This experiment shows that the computation ideality of Ara2 is related to the ratio between the application vector length and the number of lanes and that a vector is \textit{short} or \textit{long} only depending on the vector architecture on which will be processed.
On the right, the two heat plots visually remark the same concept, as entries on the same diagonal (same byte-per-lane ratio) have similar colors, \ie similar performance. 

Both analyzed kernels show a slight performance regression for a constant byte-per-lane ratio. In the case of the \texttt{dotproduct}, the performance decreases when increasing the number of lanes because of the vector reduction instruction, whose latency depends on the number of lanes during the inter-lane phase. On the other hand, \texttt{fmatmul} is computed on square matrices, and when the matrix size increases, the arithmetic intensity increases as well. Moreover, the performance drop of the \texttt{2L} configuration at 16 and 32 Byte/Lane testifies that when the number of elements is too low, the setup time of the kernel is less amortized.

\subsection{Baseline System's Performance}
\label{ssec:baseline}

\begin{figure}
    \centering
    \includegraphics[width=\linewidth]{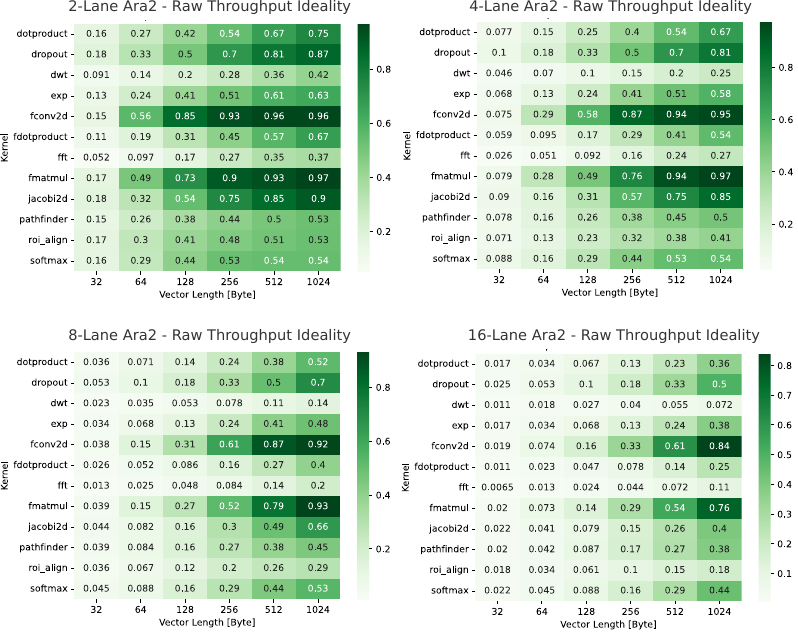}
    \caption{Performance evaluation of the system with different configurations and number of lanes on different kernels and vector lengths.}
    \label{fig:system-perf}
\end{figure}

\Cref{fig:system-perf} presents the performance ideality of the baseline system across all benchmarks in our pool, varying the application vector length. The darker the green shades in the plot, the closer the system performance is to the ideal. 
As noticed in \Cref{ssec:lanes_vec_len}, the ratio between the vector length and the number of lanes significantly affects performance, as shown by the 2-lane design reaching \revmod{over}{almost} 70\% of average performance ideality with a 1KiB vector (512 Byte/lane), while the 16-Lane system, with the same vector, would reach only 33\% (64 Byte/lane).
This is apparent in the heatmaps, where the darker colors shift right by one position every time the lanes double.

\texttt{fft}, \texttt{dwt}, \texttt{softmax}, and \texttt{pathfinder} perform below average, even for high ratios of Byte/lanes, with \texttt{softmax} suffering from the floating-point division data-dependent latency and the large setup time for preloading the approximation function coefficients of the software-emulated exponentiation. \texttt{dwt} is slowed down by misaligned strided memory accesses, and \texttt{fft} by the indexed stores at the end of the program.
However, even considering these slower kernels, the system achieves, on average, 50\% of its raw throughput ideality on all the kernels and configurations starting from 128 Byte/Lane.

The Ara2 system achieves higher performance at lower byte-per-lane ratios with the most crucial kernels like \texttt{fconv2d} and \texttt{fmatmul}, reaching over 95\% of its maximum theoretical throughput starting from 128 Byte/Lane, or 75\% from 64 Byte/Lane.

\begin{figure*}
    \centering
    \includegraphics[width=\linewidth]{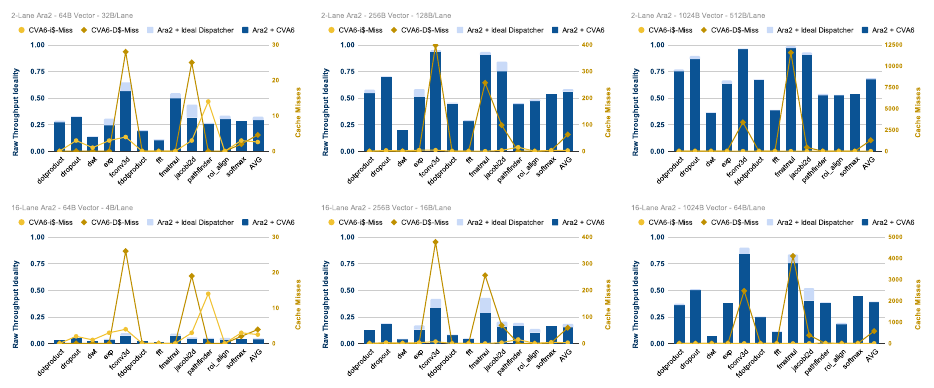}
    \caption{Scalar cache misses and performance gain when CVA6 is replaced by an ideal dispatcher, varying the number of lanes in the Ara2 system and the number of elements in the vector, measured in bytes.}
    \label{fig:cachemiss}
\end{figure*}

\subsection{Ideal Dispatcher System's Performance}
Replacing the scalar core (CVA6) with an ideal dispatcher idealizes the scalar core, its memory system, and the non-vectorial code. 
In this way, we can trace an upper bound to the available performance achievable through the Ara2 system alone, abstracted from the scalar subsystem.
With an ideal dispatcher, the performance is no longer limited by the issue-rate limitation described in \cite{Ara2020}, and CVA6 cannot interfere with Ara's memory transfers.

\Cref{fig:cachemiss} shows, for the 2- and 16-Lane system, the raw throughput ideality gain we get by replacing CVA6 with an ideal dispatcher. We report the plots for 64, 256, and 1024 elements.
The yellow lines track the number of scalar L1I-cache and L1D-Cache misses during the vector program computation, from the first vector instruction to the last one.

When the number of Bytes per lane is lower than 8 (e.g., 16-Lane Ara2 - 64B), the effective number of lanes is reduced for 64-bit data kernels since there are not enough elements to fill all the lanes, and the real maximum performance achievable is the maximum performance multiplied by the number of elements per lane. In this condition, the performance is not limited by CVA6.

When the number of Bytes per lane is lower than 64 (2-Lane Ara2 - 64B and 16-Lane Ara2 - 256B), the effective number of banks used in each lane is reduced from eight to the number of elements per lane for the 64-bit data kernels, especially if we do not implement a Barber Pole's VRF layout. For example, if the vector length is 128B and the system has 16 lanes, only one bank per lane is effectively used; all the read and write operations will target the same bank in each lane, increasing the probability of conflicts.
Since multiple \gls{PE}s experience a stall when they try to access the same bank simultaneously, and the conflict probability grows when we reduce the number of banks, these configurations are slowed down by more bank stalls. 
The lower the byte-per-lane ratio, the more impact a bank stall will have on the final runtime.

In this region, we see that some kernels moderately benefit from the ideal dispatcher. This also happens when the number of elements grows until the number of byte-per-lane becomes sufficiently long to peak performance, and the non-idealities of the scalar core and memory system are almost completely amortized by Ara\revadd{2}. 

\begin{figure}
    \centering
    \includegraphics[width=0.9\columnwidth]{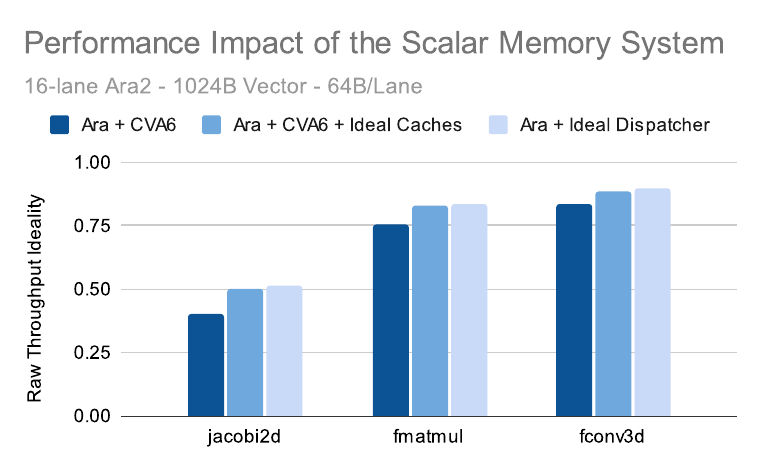}
    \caption{Performance comparison of a 16-Lane System with 64B/Lane among three different configurations: a) Baseline System, b) Baseline System with ideal CVA6 data-cache, c) System with ideal dispatcher.}
    \label{fig:large-caches}
\end{figure}

\texttt{fmatmul}, \texttt{fconv2d}, and \texttt{jacobi2d} greatly benefit from the ideal dispatcher, showing a correlation between the D-Cache misses and their performance increase when CVA6 is replaced with an ideal dispatcher. 
To prove the correlation, we simulate the three kernels on a 16-Lane System with 128 elements (1024 Byte), idealizing the cache system so that it always hits. \Cref{fig:large-caches} shows the results. With an ideal cache system, the three kernels virtually perform as when executed by the ideal-dispatcher system, which shows the importance of the scalar memory system for kernels that heavily rely on operand forwarding between the scalar and the vector core.

\revadd{Since Ara2's memory coherence logic invalidates a whole cache set per address index, the cache placement policy matters, especially for problems whose data that transit through the scalar core can be fully contained in the cache, like \texttt{fconv2d}. In this case, the larger the set (i.e., higher associativity), the more lines can be (unnecessarily) invalidated.}
\revadd{\texttt{fmatmul}, instead, requires the whole matrix A to pass through the L1 scalar d-cache. Since a 128x128 matrix does not fit the cache, the invalidation system generates mandatory capacity misses. In this case, larger cache lines reduce the number of misses since more elements are cached with every refill request.}

\revadd{Optimizing the memory hierarchy is out of the scope of this work; however, we note that a more intrusive tag-based invalidation logic would help solve the unnecessary invalidations that occur with smaller problems, and increased AXI widths for cache refills would lower the miss penalty and help all the kernels. Also, hardware or software cache prefetching would be useful to mitigate the effects of cache misses for problems with relatively low byte-per-lane ratios.}

\revmod{Instead}{Finally}, \texttt{exp} and \texttt{roi-align} slightly benefit from the ideal dispatcher because of the non-negligible amount of housekeeping scalar code in the program.

\subsection{Further Performance Considerations}
In the following paragraphs, we discuss the effect of further system/level and microarchitectural modifications used in vector processors that can impact performance and identify the main performance drivers that can boost the system's performance with low Byte/Lane ratios.

\subsubsection{VRF Layout}
The current instance of Ara2 does not implement Barber's Pole VRF layout, which was originally introduced to avoid the initial stalls that every vector instruction with at least two source registers would experience \cite{Ara2020}. These stalls are especially critical for short-vector applications where each vector instruction lasts for fewer cycles. 

Barber's Pole does not always reduce the number of runtime stalls, though. Highly regular applications such as \texttt{fmatmul} suffer more stalls when implementing this byte layout and dealing with vectors that are sufficiently long to amortize the initial stalls.
Without Barber's Pole, when an application is highly regular, and the vector processor receives enough computational vector instructions to keep its operand requesters busy, the \gls{VRF} experiences a regular access pattern during execution, especially when the vector length is multiple of the number of banks in a lane. 
After the first stalls, the source operand requesters make their requests in an interleaved fashion, once per cycle, without conflicting anymore. When an instruction is over, and the next one starts, the pattern is preserved since the previous instruction issued its last request to the last \gls{VRF} bank, while the next instruction will ask an operand to bank 0 one cycle later.
Instead, with Barber's Pole layout, the next instruction will likely start its requests from a different bank that depends on the new source register. This uncertainty perturbates the bank access pattern and creates new stalls.

\begin{figure}
    \centering
    \includegraphics[width=0.9\linewidth]{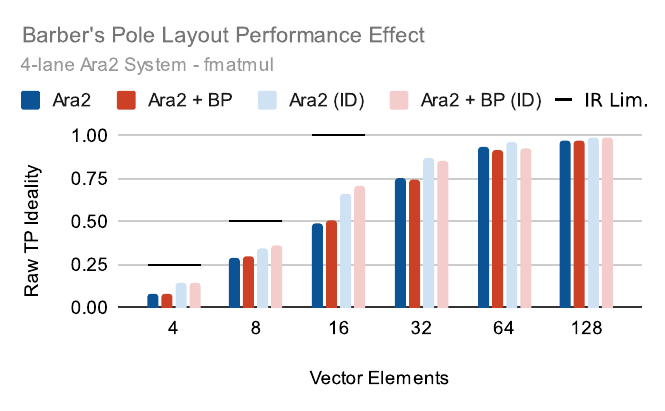}
    \vspace{-5mm}\caption{Barber's Pole layout performance effects during a matrix multiplication (8 Byte/element). Small benefits exist for shorter vectors, while medium-long vectors suffer from it.}
    \label{fig:bpole}
\end{figure}

\Cref{fig:bpole} shows this effect. We report the baseline system performance with and without Barber's Pole for \texttt{fmatmul}, varying the number of elements. We also report the same comparison but with CVA6 replaced by the ideal dispatcher. Since the trends are similar, we can focus on the baseline system. 
Barber's Pole has a positive effect on performance up to 16 double-precision elements (32 Byte per lane), as it increases the number of effective banks used in each lane. 
From 32 elements (64 Byte/lane, i.e., the effective number of banks is the same as the physical number of banks), the performance is decreased due to the newly introduced conflicts. The longer the vector, the less frequent the introduced stalls happen, as the fetch from the same source registers is not perturbed by Barber's Pole.
Barber's Pole effect on performance is mainly driven by the number of Bytes per lane; therefore, shorter data widths will experience the negative effect of this byte layout at higher vector lengths.
Since Barber's Pole layout increases the complexity of the hardware that calculates the \gls{VRF} addresses and can hurt performance when handling long vectors, we decided not to implement it in Ara2.

\subsubsection{Main Performance Drivers}

To fully characterize Ara2's performance, we studied its behavior when further optimized with modifications that can highly impact its \gls{PPA} metrics. 
To amortize the setup time of the internal pipeline when the byte-per-lane ratio is low, Ara2 needs to have larger buffers at each step of its pipeline. 
We artificially increased the size of the instruction buffers for each internal unit and the number AXI cut-registers from/to Ara's load-store unit. Moreover, we changed the way in which the hazards on the load unit and slide unit are handled, complicating the internal control but allowing for faster hazard resolution. To further decrease the number of stalls, another possible change would be to duplicate the window of simultaneous instructions handled within Ara2 from 8 to 16. 
All these changes impact the area and timing of a design that requires careful optimizations to balance the timing paths; therefore, they are only used now as a comparison point for more viable solutions presented in \Cref{sec:multi}.

\begin{figure}
    \centering
    \includegraphics[width=0.9\linewidth]{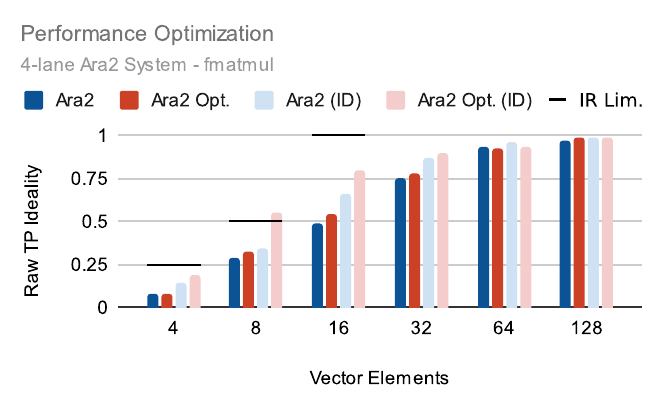}
    \vspace{-5mm}\caption{Performance analysis with further Vector Processor streamlining when executing a matrix multiplication (8 Byte/element). 
    The issue-rate limitation is a hard limit only for the system with CVA6 (opaque lines).}
    \label{fig:araopt}
\end{figure}

\Cref{fig:araopt} shows the performance of the optimized system compared to the baseline. We also report the system, in both configurations, attached to an ideal dispatcher. The black lines represent the issue-rate limitation, i.e., the intrinsic performance limitation due to the non-ideal rate at which the scalar core issues the vector instructions to Ara\revadd{2}.
We can see that the optimized system with the ideal dispatcher is not limited by the issue-rate limitation and gets a non-negligible performance boost for a vector 32, 64, and 128 bytes long (up to 32 Byte/lane) if compared with the baseline with the ideal dispatcher and with the current system baseline. On the other hand, we can see that the current system, coupled with CVA6, suffers from multiple problems with short vectors. Namely, the fact that the housekeeping code overhead and the setup time weigh more when the runtime is shorter (smaller problem size).

\begin{figure}
    \centering
    \includegraphics[width=0.9\linewidth]{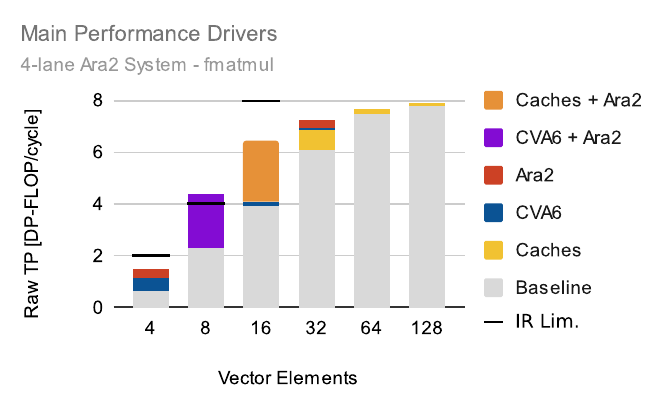}
    \caption{Main sources of inefficiency. 
    Ara2's architecture especially limits short vectors from 32 to 128 Byte (up to 32 B/lane), while caches play an important role for 128 and 256 Byte vectors. Kernel: floating-point matrix multiplication (8 Byte/element).}
    \label{fig:araineff}
\end{figure}

\Cref{fig:araineff} summarizes the results of the performance analysis, analyzing the various sources of inefficiencies for the current system. Ara's performance, when dealing with shorter vectors, is limited by multiple factors that depend on the vector length. We merged more contributions when their combined effect was way higher than their effects added (64B, 128B cases). CVA6 highly penalizes the 64B case only if Ara2 is optimized; otherwise, its inefficiencies are hidden by Ara's ones. We also want to highlight that the CVA6- and Cache-related columns are upper bounds to the system's performance. Further optimizations would require a heavy re-design of the system, as the most impactful optimizations are the merged ones (violet and orange columns). 
The effect of the Ara2 inefficiencies on performance drops below 5\% from 256B on, demonstrating the computational efficiency of the architecture for medium-long vectors.

\section{Physical implementation}
\label{sec:phy}
\begin{figure}
    \centering
    \includegraphics[width=0.9\columnwidth]{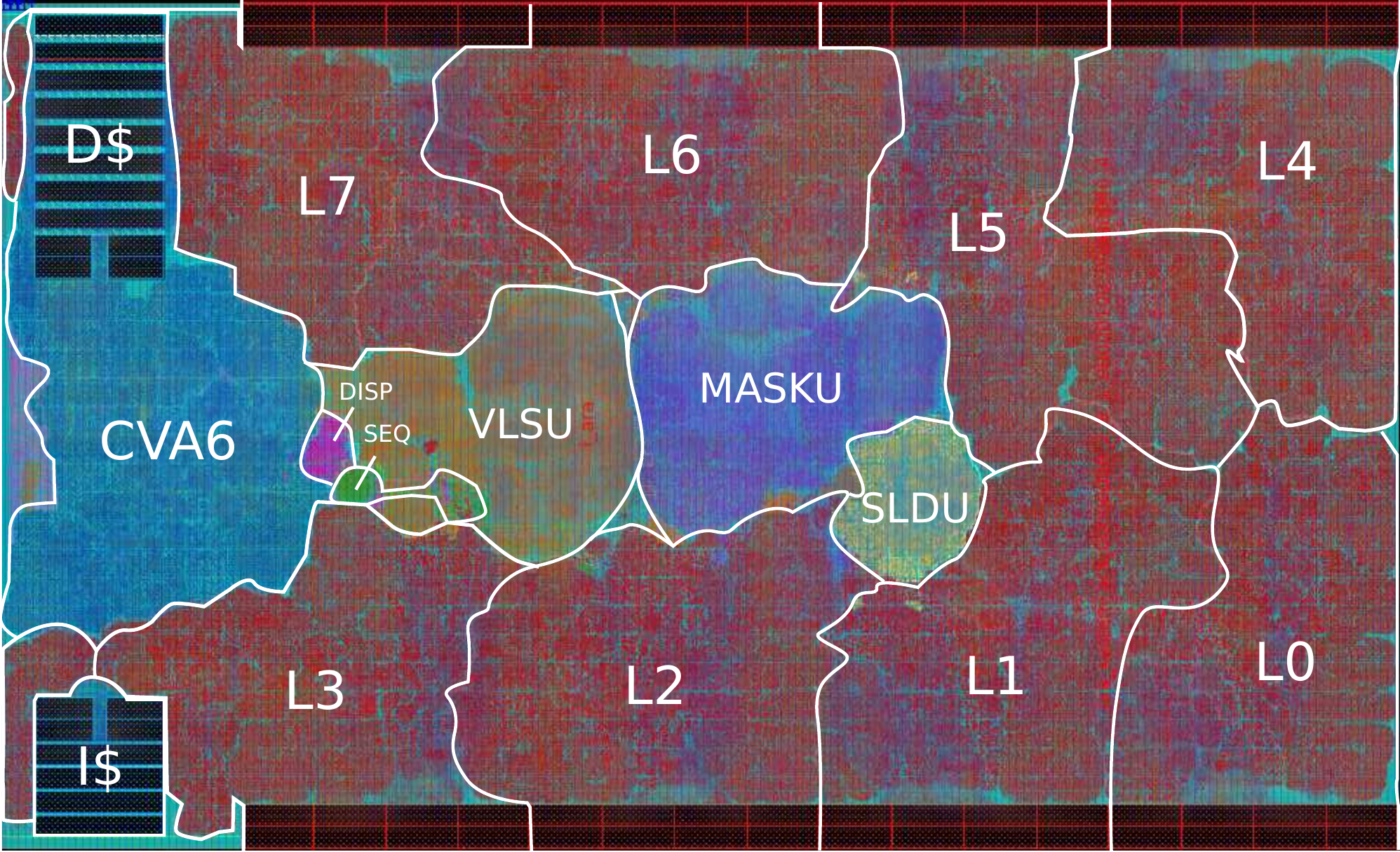}
    \caption{Floorplan of the 8-lane Ara2 implementation.}
    \label{fig:system-pnr}
\end{figure}

We summarize the system implementation results in \Cref{tab:phy}, while \Cref{tab:powbdown} presents the performance, power, and energy efficiency analysis for a \texttt{matmul} on different data types. \revadd{\Cref{fig:system-pnr} shows the 8-lane place-and-routed design.}

\begin{table}[t]
\captionsetup{font=footnotesize}
  \caption{Physical implementation metrics in 22nm FD-SOI. In brackets, the increment w.r.t. the same architecture with half the lanes.}
  \label{tab:phy}
  \centering
  \resizebox{\columnwidth}{!}{\begin{tabular}{lrrrrr}
    \toprule
    Metric (\textbf{factor}) & 2 Lanes & 4 Lanes & 8 Lanes & 16 Lanes & 16 Lanes* \\\midrule
    \textbf{SS Freq. [GHz]} & 950 & 960 (\textbf{1.0$\times$}) & 940 (\textbf{1.0$\times$}) & 750 (\textbf{0.8$\times$}) & 860 (\textbf{0.9$\times$}) \\
    \textbf{TT Freq. [GHz]} & 1.35 & 1.35 (\textbf{1.0$\times$}) & 1.35 (\textbf{1.0$\times$}) & 1.08 (\textbf{0.8$\times$}) & 1.26 (\textbf{0.9$\times$}) \\\midrule
    \color{black}\textbf{Die Area [\qty{}{\square\mm}]} & 0.59 & 0.95 (\textbf{1.6$\times$}) & 1.88 (\textbf{2.0$\times$}) & 4.47 (\textbf{2.4$\times$}) & 4.47 (\textbf{2.4$\times$}) \\
    \textbf{Cell+Macro Area [\qty{}{\kilo\gate}]} & 2291 & 3688 (\textbf{1.6$\times$}) & 6768 (\textbf{1.8$\times$}) & 14773 (\textbf{2.2$\times$}) & 12864 (\textbf{1.9$\times$}) \\
    \textbf{Macro Area [\qty{}{\kilo\gate}]} & 557 & 769 (\textbf{1.4$\times$}) & 1180 (\textbf{1.5$\times$}) & 2010 (\textbf{1.7$\times$}) & 2010 (\textbf{1.7$\times$}) \\
    \textbf{Cell Area [\qty{}{\kilo\gate}]} & 1733 & 2919 (\textbf{1.7$\times$}) & 5587 (\textbf{1.9$\times$}) & 12763 (\textbf{2.3$\times$}) & 10854 (\textbf{1.9$\times$}) \\\midrule
    \textbf{En. Eff.** [\qty{}{\flops\dp\per\watt}]} & 34.1 & 37.8 (\textbf{1.1$\times$}) & 35.7 (\textbf{0.9$\times$})& --- & 30.3 (\textbf{0.8$\times$})\\\bottomrule
    \multicolumn{6}{l}{* No fixed-point support, minimal mask unit.}\\
    \multicolumn{6}{l}{** \texttt{fmatmul} between 256x256 matrices, at typical corner}
  \end{tabular}}
\end{table}

\begin{table}[t]
\captionsetup{font=footnotesize}
  \caption{Performance, power, and energy efficiency for different kernels for a 4-lane design, at 1.35\,GHz, in typical conditions, and 2KiB vectors.}
  \label{tab:powbdown}
  \centering
  \resizebox{\linewidth}{!}{\begin{tabular}{lrrrr}
    \toprule
    Program & Elements & Power [mW] & Performance [GOPS] & Efficiency [GOPS/W] \\\midrule
    \texttt{fmatmul64} & 256 & 283 & 10.7 & 37.8 \\
    \texttt{fmatmul32} & 512 & 238 & 21.4 & 90.0 \\
    \texttt{fmatmul16} & 1024 & 218 & 42.8 & 195.9 \\
    \texttt{imatmul64} & 256 & 272 & 10.4 & 38.3 \\
    \texttt{imatmul32} & 512 & 245 & 20.9 & 85.2 \\
    \texttt{imatmul16} & 1024 & 231 & 41.8 & 181.0 \\
    \texttt{imatmul8}  & 2048 & 222 & 83.5 & 376.0 \\\bottomrule
  \end{tabular}}
\end{table}

\Cref{tab:area-bdown} shows the area breakdown of the system and the cost of supporting arbitrary strides during slide operations compared with the optimized system, which supports only power-of-two strides and time-multiplexes the slide and reshuffle operations, as explained in \Cref{sec:arch}.
When scaling up, the all-to-all unoptimized (old) slide unit becomes the largest non-lane block from four lanes on, growing exponentially with the lanes and dominating the 8-lane design. The mask unit only needs to support reshuffling and generate an enable signal at bit granularity; however, no all-to-all input-to-output byte mapping is required.
The load and store units quickly increase their size when scaling up the system but at a lower pace thanks to their limited memory bandwidth ($4 \times L$ Byte/cycle), which is half of the byte throughput of the main computational units ($8 \times L$ Byte/cycle) and needs $\frac{L^2}{2}$ connections. Between these two units, the load unit is the most problematic since its interconnect is duplicated to connect the input AXI memory bytes to the two-entry result queue. The store unit, instead, connects the input register to the AXI bytes directly.
The optimized slide unit shows a linear increase, with an area reduction of 83\% with respect to the unoptimized one. The greater reduction in area w.r.t. the predicted one can be explained by the diminished routing density that was effectively dominating the unit area.
Looking at the 16-lane design with the new slide unit, we see the mask unit and load unit exploding in size and hindering the system's scalability.

\textit{\revadd{Key insights:}}
\revadd{Designing an \gls{RVV} vector processor able to scale up to 16 lanes implies taking into account physical implementation constraints in the micro-architecture specification and design. Increasing the Byte/Lane ratio provides diminishing performance returns that eventually peak. After that point, increasing the \gls{VRF} size to allow for higher Byte/Lane ratios is no longer convenient. Indeed, we reduced the \gls{VRF} from \cite{Ara2020} by 4 without significantly impacting performance, as shown in \Cref{sec:perf}.}

\revadd{Supporting multiple element widths on wide datapaths makes the interconnects very wire-intensive, and feasibility can only be explored through full placement and routing. 
The old \gls{SLDU} interconnect area scaled up by $5\times$ when doubling the lanes from 8 to 16. With dedicated optimizations, its area now consistently scales by just {\raise.17ex\hbox{$\scriptstyle\sim$}}$2\times$, allowing the architecture to scale up further.}

\revadd{Ara2 implements a physical \gls{VRF} byte layout that eases mixed-width operations but requires reshuffling, as opposed to designs such as \cite{Minervini2022}. The new \gls{SLDU} demonstrates that the reshuffling logic can be implemented with negligible impact.}

\revadd{The memory protocol influences the complexity of the memory-to-\gls{VRF} interconnects. AXI can return misaligned elements on the bus, even if the first requested byte (variable position) maps to a fixed location (the first \gls{VRF} byte).
Indeed, the \gls{VLSU} complexity scales superlinearly with the number of lanes by $3\times$, $4.7\times$, and $6.1\times$ because of this interconnect and will be the target of future optimizations.}

\begin{table}[t]
\captionsetup{font=footnotesize}
  \caption{Ara2's area breakdown and scale-up behavior. In brackets, the area scaling factor w.r.t. the same block in Ara2 with half the lanes. Thanks to the optimization of the \gls{SLDU}, the \gls{MASKU} and \gls{VLDU} are the main units whose area skyrockets during the system upscaling. When we remove most of the shuffling functions from the \gls{MASKU}, its area is reduced by 60\% in a 16-lane design.}
  \label{tab:area-bdown}
  \centering
  \resizebox{\columnwidth}{!}{\begin{tabular}{rrrrrr}
    \toprule
    Area [kGE] (\textbf{factor}) & 2 Lanes & 4 Lanes & 8 Lanes & 16 Lanes & 16 Lanes*\\\midrule
    \textbf{CVA6} & 894 & 896 ($\mathbf{1.0 \times}$) & 906 ($\mathbf{1.0 \times}$) & 904 ($\mathbf{1.0 \times}$) & 904 ($\mathbf{1.0 \times}$) \\
    \textbf{Lane} & 612 & 617 ($\mathbf{1.0 \times}$) & 626 ($\mathbf{1.0 \times}$) & 628 ($\mathbf{1.0 \times}$) & 573 ($\mathbf{0.9 \times}$) \\
    \textbf{Dispatcher} & 16 & 17 ($\mathbf{1.1 \times}$) & 19 ($\mathbf{1.1 \times}$) & 23 ($\mathbf{1.2 \times}$) & 20 ($\mathbf{1.1 \times}$) \\
    \textbf{Sequencer} & 14 & 15 ($\mathbf{1.1 \times}$) & 17 ($\mathbf{1.2 \times}$) & 29 ($\mathbf{1.6 \times}$) & 29 ($\mathbf{1.6 \times}$) \\
    \textbf{MASKU} & 38 & 97 ($\mathbf{2.5 \times}$) & 300 ($\mathbf{3.1 \times}$) & 1105 ($\mathbf{3.7 \times}$) & 442 ($\mathbf{1.5 \times}$) \\
    \textbf{ADDRGEN} & 35 & 36 ($\mathbf{1.0 \times}$) & 44 ($\mathbf{1.2 \times}$) & 59 ($\mathbf{1.3 \times}$) & 60 ($\mathbf{1.4 \times}$) \\
    \textbf{VLDU} & 15 & 45 ($\mathbf{3.0 \times}$) & 212 ($\mathbf{4.7 \times}$) & 1286 ($\mathbf{6.1 \times}$) & 1135 ($\mathbf{5.4 \times}$) \\
    \textbf{VSTU} & 8 & 21 ($\mathbf{2.8 \times}$) & 64 ($\mathbf{3.1 \times}$) & 332 ($\mathbf{5.2 \times}$) & 342 ($\mathbf{5.3 \times}$) \\
    \textbf{New SLDU} & 24 & 48 ($\mathbf{2.0 \times}$) & 94 ($\mathbf{2.0 \times}$) & 196 ($\mathbf{2.1 \times}$) & 190 ($\mathbf{2.1 \times}$) \\\midrule
    \textbf{Old SLDU} & 39 & 131 ($\mathbf{3.4 \times}$) & 577 ($\mathbf{4.4 \times}$) & 2900 ($\mathbf{5.0 \times}$) & 2860 ($\mathbf{5.0 \times}$) \\\bottomrule
    \multicolumn{6}{l}{* No fixed-point support, minimal mask unit.}\\
  \end{tabular}}
\end{table}

\section{Multi-Core Analysis}

\begin{figure}
    \centering
    \includegraphics[width=\columnwidth]{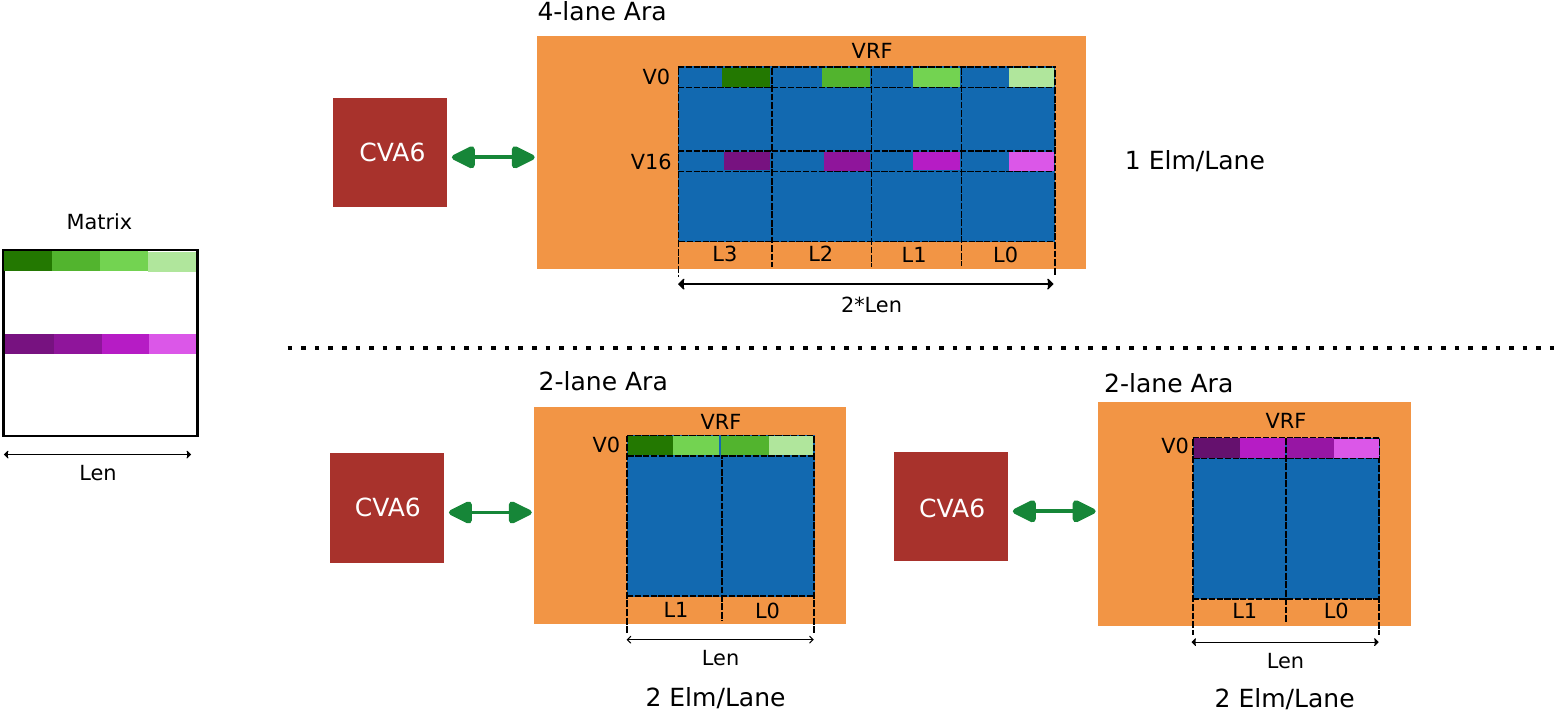}
    \caption{Comparison of a single-core 4-lane Ara2 system architecture and a two-core architecture made of 2-lane Ara2 processors. Both configurations have 4 \glspl{FPU} in total. \revmod{The vector processor can vectorize in}{Vector processors exploit} one dimension only, while \revmod{the}{a} multi-core system can parallelize on multiple dimensions. This allows the multi-core system to reach a higher Byte/Lane ratio when the vectorizable dimension is shorter than the size of a vector register and the application exposes an additional dimension.}
    \label{fig:mc-theory}
\end{figure}

\label{sec:multi}

In this section, we analyze the performance and energy efficiency trade-off for different system configurations (single-core and multi-core) as a function of the application's vector length. The analysis from \Cref{sec:phy} suggests that 16 lanes are a hard limit due to congestion and routing challenges in the all-to-all units of the vector processor. Moreover, a single-core vector processor needs to buffer longer vectors exposed by the application on a single dimension to scale up performance and efficiency when its number of lanes increases, by keeping the same byte-per-lane ratio.  However, not all applications expose long vectors on one dimension. A multi-core system offers flexibility in exploiting different parallelization dimensions at the cost of synchronization and memory-transfer overheads, as reported in \Cref{fig:mc-theory}. The experiment includes four systems with the same number of \glspl{FPU} (16), ranging from a single-core 16-lane Ara2 system to a multi-core system consisting of eight 2-lane Ara2 systems coupled with eight CVA6 instances. The results, depicted in Figures \ref{fig:issue-rate}, \ref{fig:mc-tp}, and \ref{fig:mc-ee}, showcase the raw throughput (FLOP/Cycle), real throughput (FLOPS), and energy efficiency (FLOPS/W) of the various system configurations.

\subsection{Performance}

The \textit{raw throughput} allows for evaluating the system's performance abstracted from the physical implementation.
For medium/short vectors (8, 16, 32, 64 64-bit elements), multi-core systems composed of smaller instances outperform those with fewer but larger vector cores. This is due to the exploitation of another dimension of parallelization, allowing for a byte-per-lane ratio increase. In contrast, as the problem size increases, the \glspl{FPU} are fully utilized even with the maximum number of lanes. The dual-core 8-lane and single-core 16-lane systems surpass the others at vector lengths of 128 and 256 elements, respectively, as the synchronization overhead and pressure on the memory system decrease with fewer cores. 
Extremely small problem sizes (8x8x8) suffer from the program setup time because of their lower arithmetic intensity and shorter runtime.

\begin{figure*}
\centering
\begin{minipage}[c]{0.3\textwidth}
    \centering
    \includegraphics[scale=0.40]{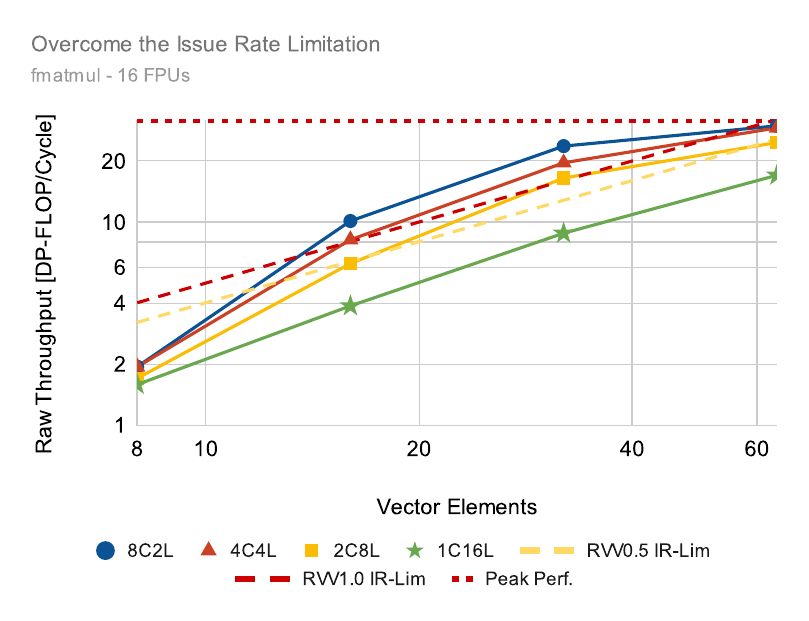}
    \vspace{-2mm}\captionof{figure}{A multi-core implementation helps overcome the issue-rate limitation for small vectors (\texttt{fmatmul}).}
    \label{fig:issue-rate}
\end{minipage}\hspace{2mm}
\begin{minipage}[c]{0.3\textwidth}
   \centering
    \includegraphics[scale=0.40]{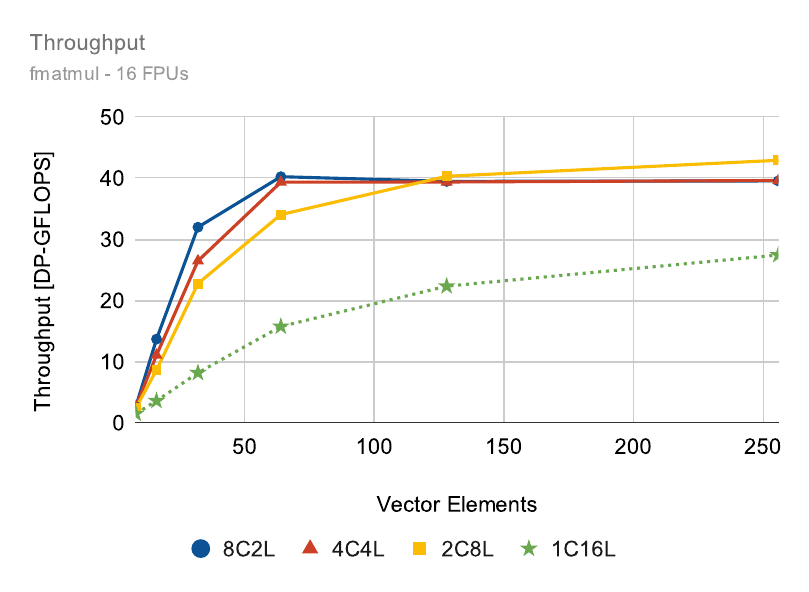}
    \vspace{-2mm}\captionof{figure}{Multi- and single-core throughput for different 16-\gls{FPU} configurations of the Ara2 system in typical conditions (\texttt{fmatmul}).}
    \label{fig:mc-tp}
\end{minipage}\hspace{2mm}
\begin{minipage}[c]{0.3\textwidth}
    \centering
    \includegraphics[scale=0.40]{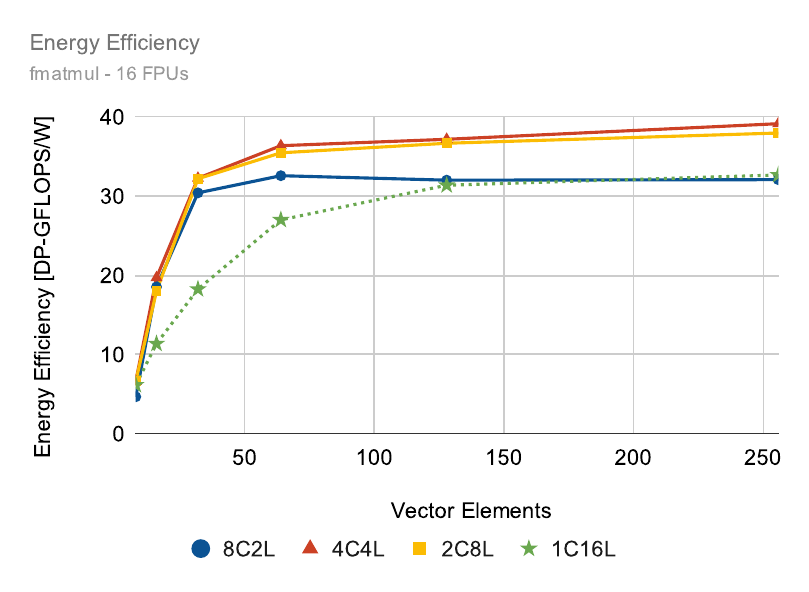}
    \vspace{-2mm}\caption{Multi- and single-core energy efficiency for different 16-\gls{FPU} configurations of the Ara2 system in typical conditions (\texttt{fmatmul}).}
    \label{fig:mc-ee}
\end{minipage}
\end{figure*}

\textit{Issue rate limitation: } 
Thanks to the \gls{RVV} specification update, the number of assembly instructions in the main loop of a \texttt{matmul} drops from four to three, bringing the issue rate of \texttt{vfmacc} instructions from five to four cycles per \texttt{vfmacc} (the scalar coefficient can now be forwarded with the \texttt{vfmacc}). 
As shown in \Cref{fig:issue-rate}, this new implicit scalar forwarding contributes to pushing the issue-rate-limitation line \cite{Ara2020} to the left.
However, the non-ideal issue rate of CVA6 still limits performance for medium-short vectors.

The multi-core vector architecture helps overcome the issue-rate limitation.
While the single-core 16-lane Ara2 system cannot theoretically go beyond 16 DP-GFLOP/cycle when operating on 32x32x32 matrices, all the multi-core instances with 16 \glspl{FPU} exceed this value, with the 8-core 2-lane system reaching 23.6 DP-GFLOP/cycle.
This happens since the maximum performance of a 2-lane Ara2 system is lower than the issue-rate limitation. Intuitively, having more scalar cores makes the total issue rate increase, too. 
Also, a multi-core design with smaller Ara2 instances behaves better than a single-core larger Ara2 with the same number of \glspl{FPU} coupled with an ideal scalar core and memory subsystem, even if the latter system is not affected by the issue-rate limitation. This result is shown in \Cref{fig:issue-rate}.

The previous analysis is valid if the application exposes two dimensions of parallelization. The applications that expose only one parallelizable dimension will always perform better on a single-core larger implementation, as the multi-core architecture will not help increase the byte-per-lane ratio and will pay the synchronization price.

As shown in \Cref{fig:mc-tp}, the overall picture of the system's performance does not change when we take into consideration the real throughput of the various systems by considering the maximum frequencies of the physical implementations for the designs up to 8 lanes. The 16-lane system experiences a non-negligible frequency drop that penalizes the whole computation, which becomes slower than all the other designs for all the different vector lengths.

\subsection{Energy efficiency}

With the multi-core system, we waste the energy of the replicated scalar cores and their memory transfers. 
This fact is counter-balanced by the different energy efficiencies reached by the physical implementations of the systems with a different lane count. 
As shown in \Cref{sec:phy}, the 4-lane design is the most efficient system, followed by the 8-lane, 2-lane, and 16-lane ones.
The multi-core overhead and the different efficiencies of the implementations allow the 4-core 4-lane design to be the most efficient one, reaching more than 39 DP-GFLOPS/W on a 256x256x256 problem. The 2-core 8-lane system immediately follows with almost 38 DP-GFLOPS/W on the same problem, thanks to a lower multi-core related power overhead.  
As expected, the 8-core 2-lane system shows an efficiency from 5\% to 18\% lower than the 4-core, suffering both from the lower efficiency of the single 2-lane Ara2 systems and the maximum number of cores in the system, with the related overhead.
The 16-lane system suffers from the specification update and the additional all-to-all connected unit that degrades the overall energy efficiency. Nevertheless, it overtakes the 8-core 2-lane system on the largest problem size.

\begin{figure*}
\centering
\begin{minipage}[c]{0.3\textwidth}
    \centering
    \includegraphics[width=\columnwidth]{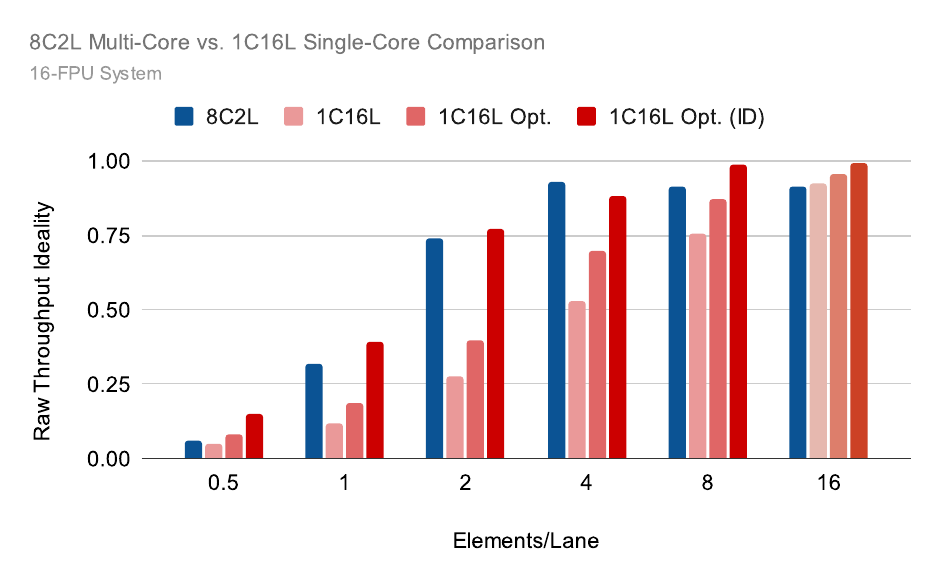}
    \captionof{figure}{Performance comparison of multi- and single-core 16-FPU Systems (\texttt{fmatmul}). In the last column of each set, CVA6 is replaced by the Ideal Dispatcher.}
    \label{fig:mc-16fpu}
\end{minipage}\hspace{2mm}
\begin{minipage}[c]{0.3\textwidth}
    \centering
    \includegraphics[scale=0.45]{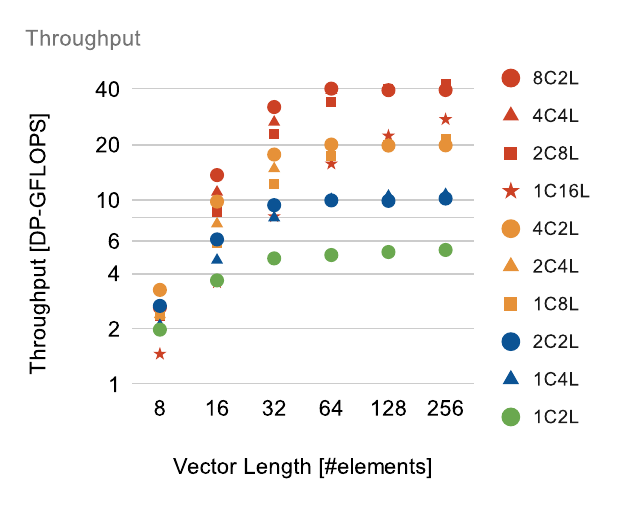}
    \captionof{figure}{Throughput comparison of single- and multi-core vector architectures with 2, 4, 8, and 16 \glspl{FPU} (\texttt{fmatmul}).}
    \label{fig:mc-ftp}
\end{minipage}\hspace{2mm}
\begin{minipage}[c]{0.3\textwidth}
    \centering
    \includegraphics[scale=0.45]{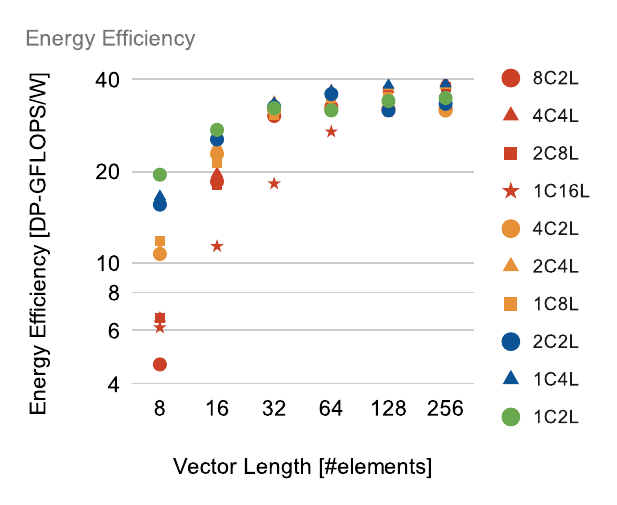}
    \captionof{figure}{Energy Efficiency comparison of single- and multi-core vector architectures with 2, 4, 8, and 16 \glspl{FPU} (\texttt{fmatmul}).}
    \label{fig:mc-feff}
\end{minipage}
\end{figure*}

In Figures \ref{fig:mc-ftp} and \ref{fig:mc-feff}, we present a log-log summary of the previous diagrams. Colors indicate the number of \glspl{FPU}, and shapes indicate the number of lanes of each Ara2 instance that makes up the multi-core system. The more complex the shape, the higher the number of lanes per core.

The larger systems with fewer cores take over the configurations with smaller cores (simple shapes on the plot) when the number of \glspl{FPU} is constant (same color on the plot) when the vector length is increased, meaning that having more lanes in the single instances is more beneficial with longer vectors, as we noticed before.

The plots show that for smaller vectors, many smaller vector processors are better than a large one that cannot effectively exploit all the dimensions of parallelism present in the application. If we take this to the limit, we see that with extremely short vectors (i.e., 1 or 2 elements), having a vector processor does not make sense, and a multi-core CVA6 scalar system would be more performant and efficient.

\begin{figure}
    \centering
    \includegraphics[width=0.8\columnwidth]{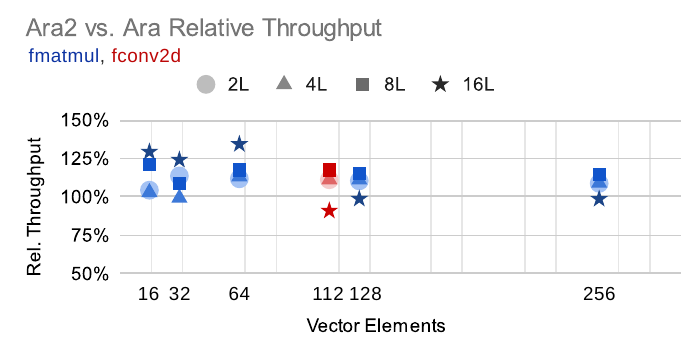}
    \caption{Performance comparison between Ara2 and Ara.}
    \label{fig:ara2-vs-ara}
\end{figure}

\section{Comparison with SoA}
\label{sec:comp}

\subsection{Performance}
As already noticed in \Cref{sec:intro}, it is not easy to compare our performance metrics with the \gls{SoA} since the benchmarking is often done on a very limited number of kernels and because the software implementation is not available. Moreover, the metrics are not always clearly defined. In \cite{Minervini2022}, we could not find the definition of the throughput metric and the lengths of the application vectors. For this reason, we assume them to be the longest possible, i.e., at least 256 DP-Elements per vector (256 Byte/Lane). The 8-lane Ara2 system is almost as performant as Vitruvius+ even with a $80 KiB/32 KiB = 2.5\times$ smaller \gls{VRF} and no register-renaming features. Indeed, it gets the same performance for \texttt{matmul} and \texttt{pathfinder}, while Ara2's FFT is $1.3 \times$ faster even for a lower byte-per-lane ratio (128 against 256).
\texttt{jacobi2d}'s performance is hard to compare as it is unclear how the authors calculated the throughput metric, getting more than what we consider the maximum throughput for that kernel (8 DP-FLOP/cycle). 
\revmod{Ara2's performance for \texttt{fmatmul, fconv2d} gains a $1.1 \times$ boost compared to [4] thanks to the frequency gain.}{\Cref{fig:ara2-vs-ara} shows the performance comparison between Ara2 and Ara on the two kernels benchmarked in \cite{Ara2020}. Ara2 is consistently faster than its counterpart (except with the 16-lane convolution) despite implementing more instructions than Ara and an up-to-date \gls{RVV} 1.0 \gls{ISA} specification. As highlighted in \Cref{sub:ppa-soa}, this was possible thanks to targeted micro-architectural optimizations that eased the datapath's timing and reduced its area while preserving performance, \eg the slide unit and the reduced \gls{VRF}, and allowed Ara2 to boost its maximum frequency, especially at higher lane counts.}
Other than that, the comparison with Hwacha done in \cite{Ara2020} is still valid, as no updated results have been published. 

The comparison with Vicuna \cite{Platzer2021} is especially hard since it is a 32-bit integer-only vector core. Its medium and fast configurations have the same computational capabilities, in terms of total throughput, of a 2-lane and 16-lane Ara2 systems. The reported utilization results are relative to two 8-bit \texttt{fmatmul} whose sizes are comparable with a $256 \times 32 \times 256$ and a $1024 \times 128 \times 1024$ 64-bit \texttt{fmatmul} problems. Our results refer to 32x32x32 and 128x128x128 matrices, and this makes the comparison unfair, as increasing the other two dimensions boosts the utilization. Nevertheless, the 2-lane Ara2 system shows virtually the same utilization. Our 16-lane system, instead, lags behind ($\mathrm{{\raise.17ex\hbox{$\scriptstyle\sim$}}30\%\ vs.\ {\raise.17ex\hbox{$\scriptstyle\sim$}}60\%}$), especially when executing the smaller problem. This is mainly due to the size of Ara's D-Cache, $16\times$ smaller than Vicuna's in the same configuration. 
Spatz \cite{Spatz2022}, an integer vector processor coupled with Snitch \cite{Zaruba2020}, is able to reach almost 70\% utilization with eight cores and 2 \glspl{IPU} per core, while Ara2 reaches 30\% for the same problem size. However, Spatz is a 32-bit integer-only vector architecture, and all the benchmark data are kept in the low latency L1 memory, which eliminates miss-related stalls.

\revadd{We report a comparison with commercial processors, even if their performance comes from fact sheets and experimental conditions are not peer-reviewed.
The Andes NX27V can compute a 32x32 32-bit \texttt{fmatmul} in {\raise.17ex\hbox{$\scriptstyle\sim$}}4800 cycles using a 32-bit vector processor with a datapath of 256-bit \cite{AndesNX27V_perf}. This is 85\% of its max performance.
A 4-lane Ara2 has the same 256-bit datapath while being a 64-bit processor. With a 64-bit 32x32 \texttt{fmatmul}, it gets 76\% of its maximum performance, which grows to 94\% with a 64x64 problem. The Intelligence X280 can reach extremely high throughput with dedicated extensions on 8-bit low-precision int and 16-bit floating-point matrix multiplications, higher than Ara2's maximum throughput. However, there is no information about performance for wider data types \cite{SiFiveX280}. As for Arm SVE, the Fujitsu A64FX reaches $90\%$ utilization on \texttt{DGEMM} \cite{a64fx_perf}, but we could not find information on the problem size.}

\revadd{Lastly, we can compare Ara2's and CVA6's performance. Comparing the vector performance against naive scalar code would unfairly favor Ara2 (e.g., 2-lane Ara2's speedup against naive scalar \texttt{fmatmul}: $70 \times$ for 128x128 matrices). 
Since a 2-lane Ara2 scores more than $90\%$ of its maximum performance with a 32x32 \texttt{fmatmul}, we can confidently say that the speedup over the scalar code is at least $2\times$ from this problem size on, as vanilla scalar cores struggle to reach peak \gls{FPU} utilization without complex superscalar out-of-order support or significant instruction extensions. For example, in \cite{Zaruba2020}, a Snitch core reaches near-peak utilization with 32x32 problems only with \revdel{both }the ad-hoc SSR and FREP \gls{ISA} extensions enabled. Without them, Snitch is an in-order scalar core, and its utilization drops below $25\%$.}
\revadd{For a direct comparison on a memory-bound kernel, we optimized a scalar dot product. 
The 2-lane Ara2 speedup over CVA6 for a floating-point dot product with 128 elements is $1.4\times$ because of the memory bandwidth limit and the \revdel{intrinsically }high \gls{FPU} latency paid at the end of the intra-lane reduction phase. Indeed, the integer dot product speedup in the same conditions is $2.2\times$.}

\subsection{Power, Performance, Area}
\label{sub:ppa-soa}
The novel Ara2 architecture boosts the design's operational frequency up to 15\% (i.e., 8 lanes) compared to Ara \cite{Ara2020}, with a critical path of {\raise.17ex\hbox{$\scriptstyle\sim$}}40 FO4 inverter delays for the 4-lane design. Its area is 2\%, 7\%, 15\%, and 37\% higher than Ara\revadd{'s} with 2, 4, 8, and 16 lanes, respectively. This is due to the newly introduced functionality, coupled with a 4$\times$ size reduction of the \gls{VRF}. The \gls{MASKU}, introduced to keep the architecture compliant with \gls{RVV}, decreases the implementation efficiency of the whole design starting from 8 lanes and a visible area, and energy-efficiency degradation hinders a further scale-up.
Further study on the \gls{MASKU} and  \gls{VLDU} is needed to reduce their footprint without hitting the system's performance. E.g., the data of the \gls{VLDU} can be aligned before entering the system, easing the routing in the unit.
On the other hand, a complete 8-lane Ara2 system (CVA6, Ara2, scalar caches) reaches virtually the same frequency and area as a Vitruvius+ instance alone without the scalar core and memories (1.4\,GHz, 1.3\,mm$^2$) \cite{Minervini2022}.

Comparing the energy efficiency across different implementations is hard due to the sensitivity that the power consumption has with respect to the input data distribution. E.g., executing the same 256x256x256 \texttt{fmatmul} kernel on our 4-lane Ara2 System can lead to energy efficiencies from 38.8 to 65 \qty{}{\giga\flops\dp\per\watt} depending on the input distribution. Using a normal distribution with mean 0 and variance 1, instead of a uniform one between [0,1), also leads to a 7.5\% power difference.
In this scenario, comparing our energy efficiency with Vitruvius+ is hard, as the paper does not report the used input number distribution, and it is not clear if the power figures consider the scalar core and memories too, as we do. Moreover, we could not reproduce the number reported in their comparison tables.
A comparison with Spatz is also nontrivial, as the architecture is a 32-bit integer only, leading to substantially lower power figures.

\subsection{RVV vs. Arm SVE}

\begin{figure}
    \centering
    \includegraphics[width=0.9\columnwidth]{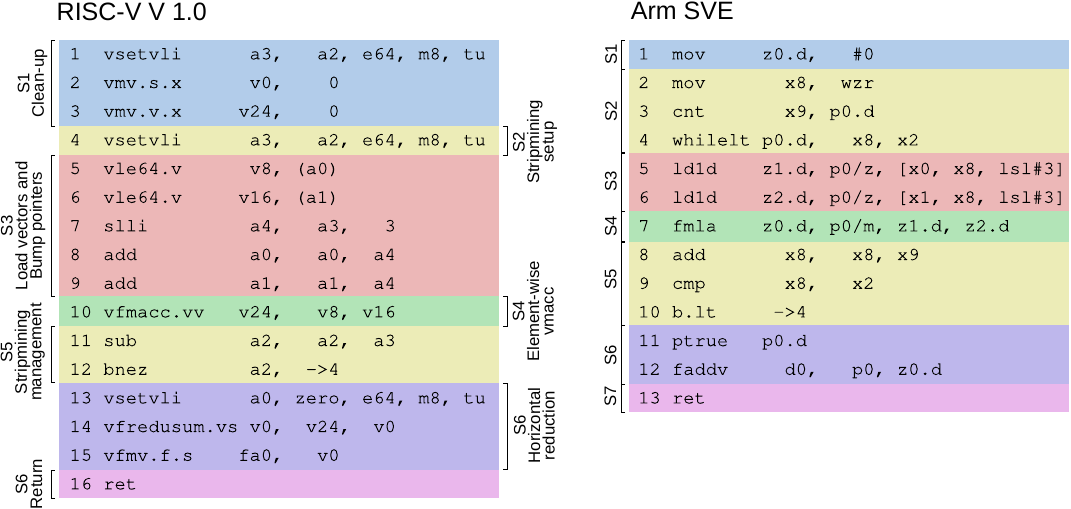}
    \caption{Comparison between RVV 1.0 and Arm SVE assembly.}
    \label{fig:rvv-vs-sve}
\end{figure}

\revadd{In \Cref{fig:rvv-vs-sve}, we report a comparison between an \gls{RVV} 1.0 and Arm SVE simplified \texttt{dotproduct} in assembly. The instruction count is $7 + 9N$ and $6 + 7N$ for RISC-V and Arm, respectively, where $N$ is the number of iterations of the strip-mining loop. Arm's complex addressing scheme allows loading a vector and bumping the address pointer in one instruction, while \gls{RVV} needs three. Also, Arm does not need to clear the scalar result register in S6. However, RISC-V can set up the strip-mining loop and get the effective vector length in just one instruction and, in this example, exploit its \texttt{bnez} compare-and-branch, which Arm splits into two instructions. Overall, Arm's slight advantage comes from its CISC-like addressing; this observation points to opportunities for further optimization to be tackled with future \gls{RVV} extensions.}

\section{Conclusion}

In this work, we presented Ara2, the first open-source \gls{RVV} 1.0 vector architecture. We implement the design with 2, 4, 8, and 16 lanes in 22nm \gls{FD-SOI} technology and reach 1.35\,GHz (for 8 and fewer lanes) thanks to a new lightweight \gls{SLDU} \revadd{datapath} able to reduce the \revdel{ number of }interconnect wiring and hardware \revadd{cost} of the all-to-all byte-connected unit by 70\%. 
For compliance with \gls{RVV}, Ara2 requires a new all-to-all connected \gls{MASKU} to support the modified \gls{VRF} layout of the mask vectors, complicating the physical implementation. This, together with the newly supported features like floating-point reductions, make the 4-lane configuration become the most efficient design point, with an \gls{SoA} energy efficiency of almost \qty{39}{\giga\flops\dp\per\watt} during a matrix multiplication. 

The flexible Ara2 architecture reaches, on average, more than 50\% of its throughput ideality starting from 128 Bytes per lane on the whole benchmark pool with every system configuration. Furthermore, with only a 64 byte-per-lane ratio, it achieves more than 75\% of its maximum throughput on the most crucial kernels like matrix multiplications and convolutions and more than 90\% from 128 Byte/Lane.
We also investigate the detrimental impact of the Barber's Pole layout on long-vector performance and thoroughly analyze the primary performance drivers of the vector architecture, including the scalar core and its memory system. 

To overcome the issue-rate performance limitation, we evaluate multi-core vector architectures,
which show that smaller vector core instances can boost the performance of applications that expose more than one dimension of parallelization (e.g., \texttt{matmul}), especially when the vectorizable dimension cannot provide high byte-per-lane ratios, with performance improvements up to 3x when compared to a single-core architecture with the same overall computation capability. This is especially beneficial when the system cannot exploit the advantages of long vectors, e.g., when the application does not expose them on a single dimension.

\ifCLASSOPTIONcompsoc
  \section*{Acknowledgments}
\else
  \section*{Acknowledgment}
\fi

This work was supported by the ETH Future Computing Laboratory
(EFCL), financed by a gift from Huawei Technologies, and by the TRISTAN (101095947) project that received funding from the HORIZON CHIPS-JU programme.

\ifCLASSOPTIONcaptionsoff
  \newpage
\fi

\bibliographystyle{IEEEtran}
\bibliography{ara,ieeetran}

\appendices

\vskip -2\baselineskip plus -1fil
\begin{IEEEbiography}[{\includegraphics[width=1in, height=1.25in, clip, keepaspectratio]{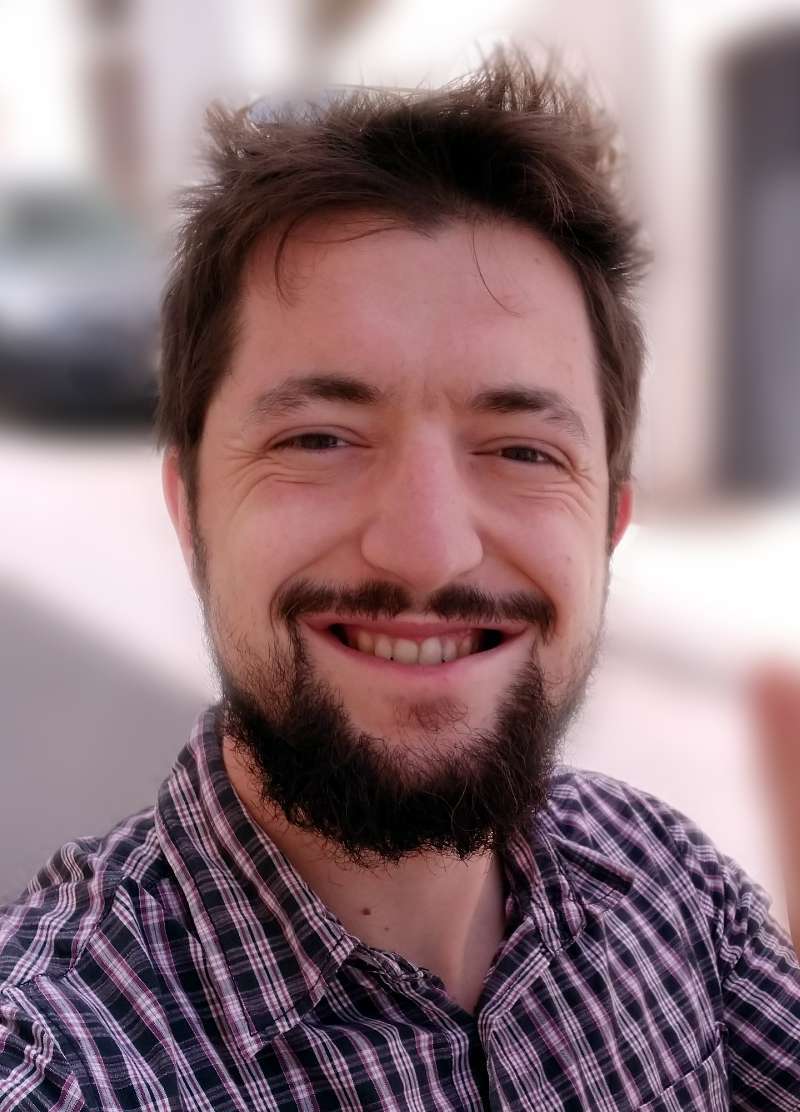}}]{Matteo Perotti}
received his M.Sc. degree in Electronic Engineering from the Polytechnic University
of Turin, Italy, in 2019. He is currently pursuing a
Ph.D. degree at the Integrated Systems Laboratory
of ETH Zurich, Switzerland, under the supervision of Prof. Luca Benini. Matteo's research interests include highly efficient computing architectures, vector processing, computation with high dynamic-range data types, and the RISC-V ecosystem in general. 
\end{IEEEbiography}
\vskip -2.5\baselineskip plus -1fil
\begin{IEEEbiography}[{\includegraphics[width=1in, height=1.25in, clip, keepaspectratio]{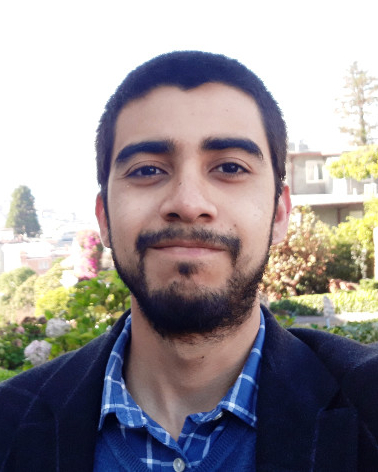}}] {Matheus Cavalcante} received his M.Sc. degree in Integrated Electronic Systems from the Grenoble Institute of Technology (Phelma) in 2018, and his Ph.D. from ETH Zurich in 2023. During his Ph.D. studies, Matheus worked with the Digital Circuits and Systems Group under the supervision of Prof. Luca Benini. Matheus’ research interests include vector processing, large-scale high-performance computer architectures, and emerging VLSI technologies.
\end{IEEEbiography}
\vskip -2.5\baselineskip plus -1fil
\begin{IEEEbiography}[{\includegraphics[width=1in,height=1.25in,clip,keepaspectratio]{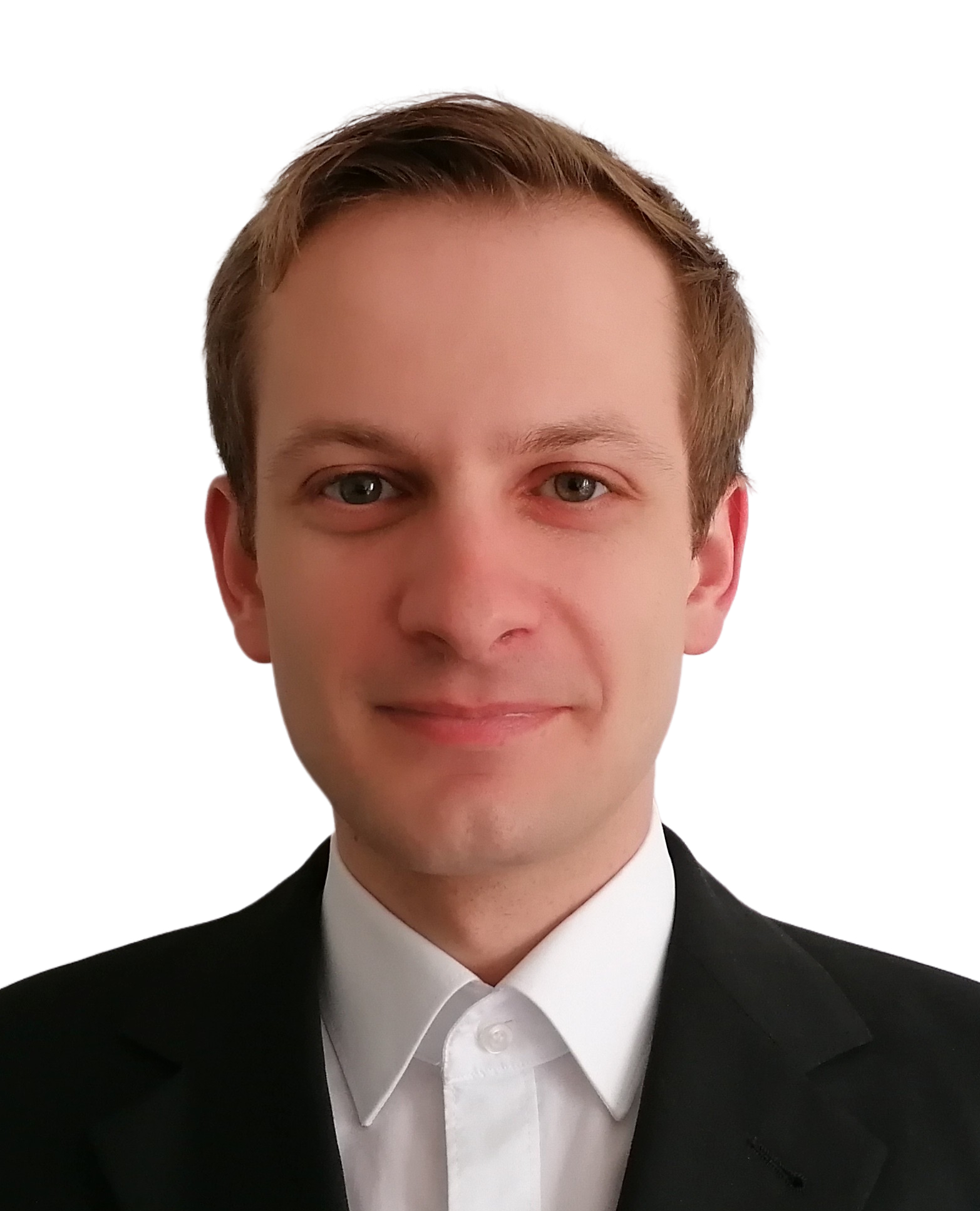}}]{Renzo Andri} received the B.Sc., M.Sc. and Ph.D. degree in Electrical Engineering and Information Technology at ETH Zurich in 2013, 2015, and 2020, respectively. Since 2020 he is working at the Huawei Research Center Zurich. His research focuses lies on energy-efficient machine learning acceleration and processor design down to full-custom IC design including hardware-algorithm codesign. In 2019, he won the IEEE TCAD Donald O. Pederson Award.
\end{IEEEbiography}
\vskip -2.5\baselineskip plus -1fil
\begin{IEEEbiography}[{\includegraphics[width=1in,height=1.25in,clip,keepaspectratio]{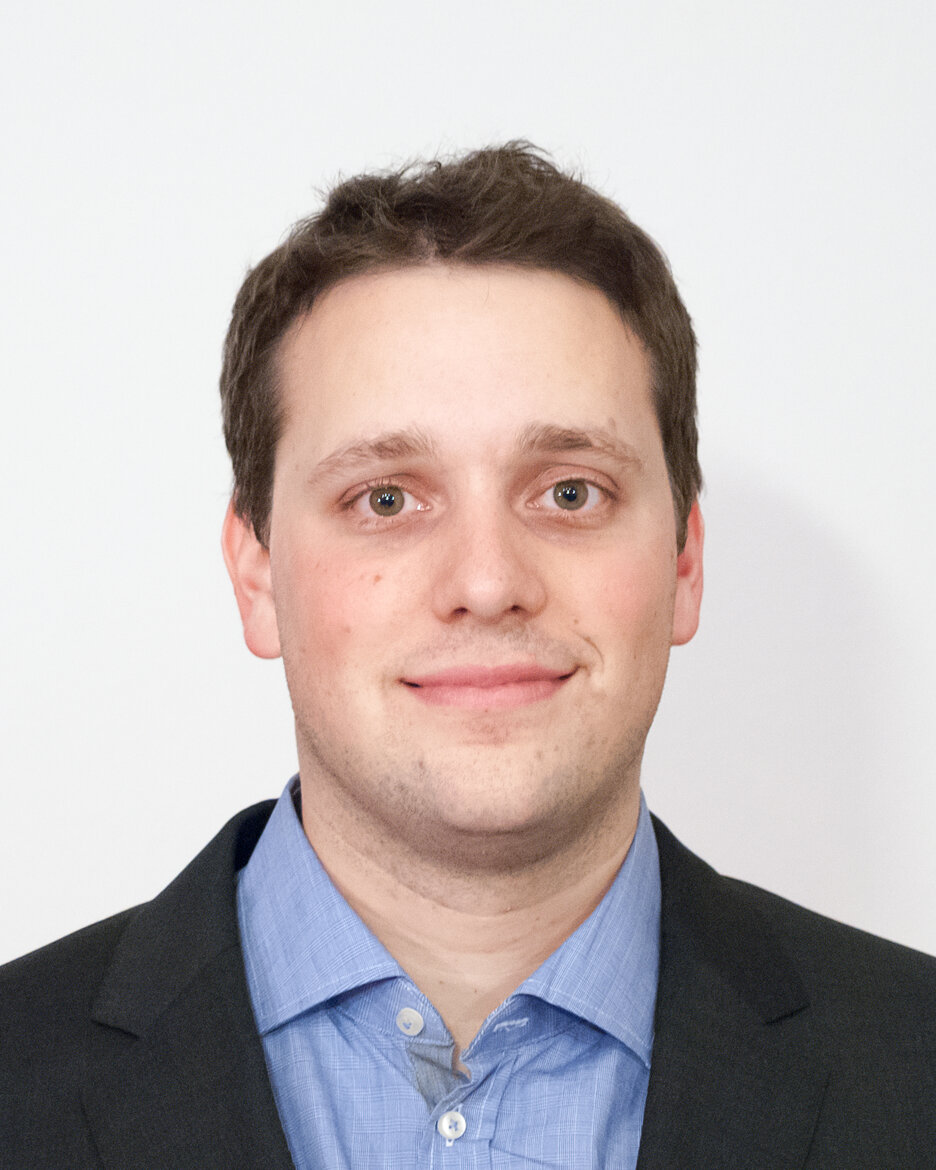}}]{Lukas Cavigelli} received the d Ph.D. degree in electrical engineering and information technology from ETH Zurich in 2019. After a year as a Postdoc at ETH Zurich, he has joined Huawei's Zurich Research Center in Spring 2020. His research interests include deep learning, computer vision, embedded systems, and low-power integrated circuit design. He has received several best paper awards including the IEEE TCAD Donald O. Pederson award in 2019.
\end{IEEEbiography}
\vskip -2.5\baselineskip plus -1fil
\begin{IEEEbiography}[{\includegraphics[width=1in, height=1.25in, clip, keepaspectratio]{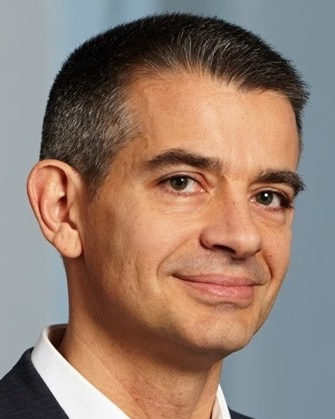}}] {Luca Benini} holds the chair of Digital Circuits and Systems at ETH Zurich and is Full Professor at the Università di Bologna. He received a Ph.D.\ from Stanford University. Dr.\ Benini's research interests are in energy-efficient parallel computing systems and ML hardware. He is a Fellow of the ACM and a member of the Academia Europaea. He received the IEEE Mac Van Valkenburg Award in 2016, the ACM/IEEE A.\ Richard Newton Award in 2020, and the IEEE E.\ McCluskey Award in 2023.
\end{IEEEbiography}

\end{document}

%% file: acronyms.tex
\newacronym[longplural={Scratchpad Memories}]{SPM}{SPM}{Scratchpad Memory}
\newacronym{ACE}{ACE}{AXI Coherent Extensions}
\newacronym{AMBA}{AMBA}{Advanced Microcontroller Bus Architecture}
\newacronym{APB}{APB}{Advanced Peripheral Bus}
\newacronym{API}{API}{Application Programming Interface}
\newacronym{ASIC}{ASIC}{Application-Specific Integrated Circuit}
\newacronym{AVX}{AVX}{Advanced Vector Extension}
\newacronym{AXI}{AXI}{Advanced eXtensible Interface}
\newacronym{BLAS}{BLAS}{Basic Linear Algebra Subprograms}
\newacronym{CHI}{CHI}{Coherent Hub Interface}
\newacronym{CMOS}{CMOS}{Complementary Metal-Oxide-Semiconductor}
\newacronym{CNN}{CNN}{Convolutional Neural Network}
\newacronym{CPU}{CPU}{Central Processing Unit}
\newacronym{CSR}{CSR}{Control and State Register}
\newacronym{CTS}{CTS}{Clock Tree Synthesis}
\newacronym{DLP}{DLP}{Data Level Parallelism}
\newacronym{DMA}{DMA}{Direct Memory Access}
\newacronym{DRAM}{DRAM}{Dynamic Random-Access Memory}
\newacronym{DSP}{DSP}{Digital Signal Processing}
\newacronym{DUT}{DUT}{Device Under Test}
\newacronym{EW}{EW}{Element Width}
\newacronym{EEW}{EEW}{Effective Element Width}
\newacronym{ECL}{ECL}{Emitter-Coupled Logic}
\newacronym{FBB}{FBB}{Forward Body-Biasing}
\newacronym{FDSOI}{FD-SOI}{Fully Depleted Silicon-on-Insulator}
\newacronym{FMA}{FMA}{Fused Multiply-Add}
\newacronym{FPGA}{FPGA}{Field-Programmable Gate Array}
\newacronym{FP}{FP}{Floating Point}
\newacronym{FPU}{FPU}{Floating Point Unit}
\newacronym{GPGPU}{GPGPU}{General-Purpose \acrlong{GPU}}
\newacronym{GPU}{GPU}{Graphics Processing Unit}
\newacronym{HDL}{HDL}{Hardware Description Language}
\newacronym{HERO}{HERO}{Heterogeneous Embedded Research Platform}
\newacronym{HPC}{HPC}{High-Performance Computing}
\newacronym{ILP}{ILP}{Instruction Level Parallelism}
\newacronym{IoT}{IoT}{Internet-of-Things}
\newacronym{IOT}{IoT}{Internet-of-Things}
\newacronym{IPC}{IPC}{Instructions Per Cycle}
\newacronym{IPU}{IPU}{Image Processing Unit}
\newacronym{ISA}{ISA}{Instruction Set Architecture}
\newacronym{LSB}{LSB}{Least Significant Bit}
\newacronym{LSU}{LSU}{Load/Store Unit}
\newacronym{LVT}{LVT}{low voltage threshold}
\newacronym{MIMD}{MIMD}{multiple instruction, multiple data}
\newacronym{MMU}{MMU}{Memory Management Unit}
\newacronym{MUL}{MUL}{multiplier}
\newacronym{ML}{ML}{Machine Learning}
\newacronym{NoC}{NoC}{network on chip}
\newacronym{MVL}{MVL}{maximum vector length}
\newacronym{NUMA}{NUMA}{non-uniform memory access}
\newacronym{NOC}{NoC}{Network-on-Chip}
\newacronym{PCIe}{PCIe}{Peripheral Component Interconnect Express}
\newacronym{PC}{PC}{Program Counter}
\newacronym{PE}{PE}{processing element}
\newacronym{PL}{PL}{Programmable Logic}
\newacronym{PMCA}{PMCA}{Programmable Manycore Accelerator}
\newacronym{PPA}{PPA}{Power, Performance, Area}
\newacronym{PSL}{PSL}{Power Service Layer}
\newacronym{PTE}{PTE}{page-table entry}
\newacronym{PTW}{PTW}{page-table walker}
\newacronym{PULP}{PULP}{Parallel Ultra Low Power}
\newacronym{RAW}{RAW}{read-after-write}
\newacronym{RBB}{RBB}{Reverse Body-Biasing}
\newacronym{ROB}{ROB}{Reorder Buffer}
\newacronym{RTL}{RTL}{Register Transfer Level}
\newacronym{RVT}{RVT}{Regular Voltage Threshold}
\newacronym{RoCC}{RoCC}{Rocket Custom Coprocessor Interface}
\newacronym{SCM}{SCM}{Storage Class Memory}
\newacronym{SEW}{SEW}{Single Element Width}
\newacronym{SIMD}{SIMD}{single instruction, multiple data}
\newacronym{SIMT}{SIMT}{single instruction, multiple thread}
\newacronym{SLDU}{SLDU}{Slide Unit}
\newacronym{SLVT}{SLVT}{super-low voltage threshold}
\newacronym{SM}{SM}{Streaming Multiprocessor}
\newacronym[longplural={Static Random-Access Memories}]{SRAM}{SRAM}{Static Random-Access Memory}
\newacronym{SSE}{SSE}{Streaming SIMD Extension}
\newacronym{SVE}{SVE}{Scalable Vector Extension}
\newacronym{TLP}{TLP}{Thread Level Parallelism}
\newacronym{TxnID}{TxnID}{Transaction ID}
\newacronym{VAC}{VAC}{Vector Access}
\newacronym{VC}{VC}{virtual channel}
\newacronym{VCONV}{VCONV}{Vector Conversion}
\newacronym{VEX}{VEX}{Vector Execute}
\newacronym{VFU}{VFU}{vector functional unit}
\newacronym{VID}{VID}{Vector Instruction Decode}
\newacronym{VIS}{VISSUE}{Vector Instruction Issue}
\newacronym{VLIW}{VLIW}{Very Long Instruction Word}
\newacronym{VLOOP}{VLOOP}{Vector Loop}
\newacronym{VLR}{VLR}{vector length register}
\newacronym{VLSU}{VLSU}{Vector Load/Store Unit}
\newacronym{VNB}{VNB}{Von Neumann Bottleneck}
\newacronym{VRF}{VRF}{Vector Register File}
\newacronym{VT}{VT}{vector thread}
\newacronym{BW}{BW}{bandwidth}
\newacronym{MASKU}{MASKU}{Mask Unit}
\newacronym{VU0.5}{VU0.5}{Vector Unit 0.5}
\newacronym{VU1.0}{VU1.0}{Vector Unit 1.0}
\newacronym{VMFPU}{VMFPU}{Vector Multiplier/Floating Point Unit}
\newacronym{VFPU}{VFPU}{Vector Floating Point Unit}
\newacronym{VDIV}{VDIV}{Vector Divider}
\newacronym{VMUL}{VMUL}{Vector Multiplier}
\newacronym{WAR}{WAR}{write-after-read}
\newacronym{WAW}{WAW}{write-after-write}
\newacronym{DCT}{DCT}{discrete cosine transform}
\newacronym{TSV}{TSV}{through-silicon via}
\newacronym{3DIC}{3D-IC}{three-dimensional integrated circuit}
\newacronym{F2F}{F2F}{face-to-face}
\newacronym{IC}{IC}{integrated circuit}
\newacronym{C4}{C4}{controlled collapse chip connection}
\newacronym{FEOL}{FEOL}{front end of the line}
\newacronym{BEOL}{BEOL}{back end of the line}
\newacronym{SLEN}{SLEN}{striping distance}
\newacronym{VSU}{VSU}{Vector Store Unit}
\newacronym{DNN}{DNN}{Deep Neural Networks}
\newacronym{AI}{AI}{Artificial Intelligence}
\newacronym{AR}{AR}{Augmented Reality}
\newacronym{SoA}{SoA}{State-of-the-Art}
\newacronym{FD-SOI}{FD-SOI}{Fully Depleted - Silicon on Insulator}
\newacronym{RVV}{RVV}{RISC-V ``V''}
\newacronym{ALU}{ALU}{Arithmetic-Logic Unit}
\newacronym{FFT}{FFT}{Fast Fourier Transform}
\newacronym{VLDU}{VLDU}{Vector Load Unit}